\documentclass[preprint2,longabstract]{aastex}
\usepackage{lscape}
\newcommand{\Msun}{\mbox{$M_{\odot}$}}

\newcommand{\jh }[1]{{#1}}

\newcommand{\Spitzer}{{\em Spitzer}}
\newcommand{\SpitzerGB}{SGBS}

\shorttitle{Spitzer GBS: V. Oph N}
\shortauthors{Hatchell et al.}

\begin{document}

\title{The {\em Spitzer} Survey of Interstellar Clouds in the Gould Belt. V.  Ophiuchus North Observed with IRAC and MIPS.}

\author{J. Hatchell}
\affil{Astrophysics Group, Physics, University of
  Exeter, Exeter, EX4 4QL, United Kingdom}
\email{hatchell@astro.ex.ac.uk}

\author{S. Terebey}
\affil{Department of Physics and Astronomy PS315, 5151 State University Drive, California State University at Los Angeles, Los Angeles, CA 90032, USA}
\email{sterebe@calstatela.edu}

\author{T. Huard}
\affil{Department of Astronomy, University of Maryland, College Park, MD 20742, USA; Columbia Astrophysics Laboratory, Columbia University, New York, NY 10027, USA}
\email{thuard@astro.umd.edu}

\author{E. E. Mamajek}
\affil{
University of Rochester, Department of Physics \& Astronomy, P.O. Box 270171, Rochester, NY, 14627-0171 USA\altaffilmark{1}}
\email{emamajek@pas.rochester.edu}

\author{L. Allen}
\affil{National Optical Astronomy Observatory, 950 N. Cherry Ave., Tucson, AZ 85719, USA}
\email{lallen@noao.edu}

\author{T.L. Bourke}
\affil{Harvard-Smithsonian Center for Astrophysics, Cambridge, MA 02138, USA }
\email{tbourke@cfa.harvard.edu}

\author{\jh{M. M. Dunham}}
\affil{\jh{Department of Astronomy, Yale University, Box 208101, New Haven, CT 06520-8101, USA}}
\email{\jh{michael.dunham@yale.edu}}

\author{R. Gutermuth}
\affil{Department of Astronomy, University of Massachusetts, Amherst, MA 01002, USA}
\email{rgutermu@astro.umass.edu}

\author{P. M. Harvey}
\affil{Astronomy Dept., University of Texas, Austin, TX 78712, USA}
\email{pmh@astro.as.utexas.edu}

\author{J. K. J{\o}rgensen}
\affil{Niels Bohr Institute and Centre for Star and Planet Formation, University of Copenhagen, Juliane Maries Vej 30, DK-2100 Copenhagen {\O}, Denmark.}
\email{jeskj@nbi.dk}

\author{B. Mer\'{i}n}
\affil{Herschel Science Centre, European Space Astronomy Centre (ESA), P.O. Box, 78, 28691 Villanueva de la Ca\~{n}ada, Madrid, Spain}
\email{Bruno.Merin@sciops.esa.int}

\author{A. Noriega-Crespo}
\affil{Science Center, MC 220-6, California Institute of Technology, Pasadena, CA 91125, USA}
\email{alberto@ipac.caltech.edu}

\and

\author{\jh{D. E. Peterson}}
\affil{\jh{Harvard-Smithsonian Center for Astrophysics, Cambridge, MA 02138, USA} }
\email{\jh{dpeterson@cfa.harvard.edu}}

%\email{} include co-authors' emails?
\altaffiltext{1}{\jh{Current address: Cerro Tololo Inter-American Observatory, Cassilla 603, La Serena, Chile}}
\clearpage

\begin{abstract}
We present \Spitzer\ IRAC (2.1 sq. deg.) and MIPS (6.5 sq. deg.) observations of star formation in the Ophiuchus North molecular clouds.  This fragmentary cloud complex lies on the edge of the Sco-Cen OB association, several degrees to the north of the well-known $\rho$~Oph star-forming region, at an approximate distance of 130~pc.  The Ophiuchus~North clouds were mapped as part of the \Spitzer\ Gould Belt project under the working name `Scorpius'.  In the regions mapped, selected to encompass all the cloud with visual extinction $A_V>3$, eleven Young Stellar Object (YSO) candidates are identified, eight from IRAC/MIPS colour-based selection and three from 2MASS~$K_S$/MIPS colours.  Adding to one source previously identified in L43 \citep{chen09}, this increases the number of YSOcs identified in Oph~N to twelve. During the selection process, four colour-based YSO candidates were rejected as probable AGB stars and one as a known galaxy.  The sources span the full range of YSOc classifications from Class~0/1 to Class~III, and starless cores are also present.   Twelve high-extinction ($A_V>10$) cores are identified with a total mass of $\sim 100$~\Msun.  These results confirm that there is little ongoing star formation in this region (instantaneous star formation efficiency $<0.34$\%)  and that the bottleneck lies in the formation of dense cores.  The influence of the nearby Upper~Sco~OB association, including the 09V~star $\zeta$~Oph, is seen in dynamical interactions and raised dust temperatures but has not enhanced levels of star formation in Ophiuchus~North.

\end{abstract}


\section{Introduction}
\label{sect:introduction}
% Spitzer GB

%Context

The Ophiuchus~North (Oph~N) molecular clouds lie $20^\circ$ above the Galactic plane in the direction of the Galactic Centre.  They are part of the same filamentary cloud complex as the well-studied Ophiuchus~L1688 and L1689 clouds, but lie several degrees to the north, on the boundary of the constellation of Ophiuchus with Scorpius.  Like the Ophiuchus~~L1688 and L1689 clouds, they are illuminated from the northwest by the Upper~Scorpius subgroup of the Sco-Cen OB association.   

The region has been studied little. Early low-resolution CO mapping \citep{degeus90,degeus92} showed the filamentary structure of the clouds in the Ophiuchus % and Scorpius
 region (mirrored in the extinction maps published by \citealt{dobashi05} and \citealt{rowlesfroebrich09}), and suggested a shock origin due to expanding shells surrounding the Upper~Sco subgroup.    A detailed study of the molecular clouds was made in $^{13}$CO by \citet{nozawa91} which gives an excellent overview of the cloud complex, and its relationship to the Ophiuchus cores, IRAS sources, and the Sco-Cen OB association.  They find that the region contains some 23 $^{13}$CO clouds containing 51 $^{13}$CO cores, with a total mass of 4000\Msun\ and typical core densities of $N_{{\mathrm H}_2} \sim 3\times 10^3\hbox{ cm}^{-3}$.  The dense cores ($N_{\mathrm{H}_2} \sim 10^4\hbox{ cm}^{-3}$) and velocity structure were subsequently followed up in C$^{18}$O by \citet{tachihara00b,tachihara00a,tachihara02}. \citet{nozawa91} identify only thirteen YSOs associated with the cores, pointing to a low star formation efficiency of 0.3\%.  

% Look in notes... 20080709, 

%Distance
 Distance estimates for the Oph~N molecular cloud complex come from its relationship with the stars in Upper Sco (US) as, from extinction, the molecular clouds lie in front of and distributed through the OB population \citep{degeus89}.  Hipparcos parallaxes place Upper Sco at a mean distance of $145\pm2$~pc \citep{dezeeuw99}, with a line-of-sight extent of $\pm 17$~pc assuming the $14^\circ$ spatial extent is reproduced in the third dimension.  This places an effective upper limit on the clouds' distances of 162~pc.  Extinction-based distance modulus estimates suggest the clouds are distributed between 80~pc (near side) and 170~pc (far side), centre 125~pc \citep{degeus89} or, slightly further away, 120~pc (near side) to 200~pc (far side) \citep{straizys84}.  These distances are consistent with the Hipparcos data.   Looking at individual stars, \citet{degeus90} suggest that the western clouds (our OphN~4,5,6  their Complex~2) lie in front of $\chi$~Oph (150~pc) and the northeastern (OphN~1/2, Complex~4) in front of $\xi$~Oph (130~pc).
%, with individual photometric distances from \citet{degeus88}.  
Recent estimates for the better-studied L1688 Ophiuchus cloud range from 120--145~pc \citep{wilking08}.
There is certainly no reason to believe that the Ophiuchus clouds all lie at the same distance, but for convenience we assume a working distance of 130~pc, which is consistent with the above estimates.

In this paper, we present mid-infrared, \Spitzer\ Space Telescope observations of the high column density regions of Ophiuchus~North.  The observations and data reduction are described in Sect.~\ref{sect:observations}.  Results, including source statistics, YSO candidates, extinction maps and large-scale emission, are given in Sect.~\ref{sect:results} with comments on individual regions in Appendix~\ref{sect:regions}.  The results are discussed in Sect.~\ref{sect:discussion} and summarised in Sect.~\ref{sect:summary}.

\section[]{Description of Observations}
\label{sect:observations}

% Table of regions with dates, program IDs

We observed Ophiuchus~North in the mid-infrared as part of the \Spitzer\ legacy program ``Gould's Belt: star formation in the solar neighbourhood'' (\SpitzerGB).  The clouds were mapped under the working name `Scorpius' (Sco) and appear under this name in the \SpitzerGB\ catalogues.  We present them here as the Ophiuchus~North (OphN) clouds in line with previous nomenclature \citep{nozawa91,tachihara00a, tachihara00b,tachihara02} which reflects their location predominantly within the constellation of Ophiuchus.  Only our regions OphN~5,6 and LDN~43 lie beyond the constellation boundary in Scorpius.

\begin{landscape}
\begin{figure}
\epsscale{.80}
%\plottwo{PS/coverage_ann_bw.eps}{PS/coverage_ann.eps}
%\plotone{PS/coverage_ann.eps}
\plotone{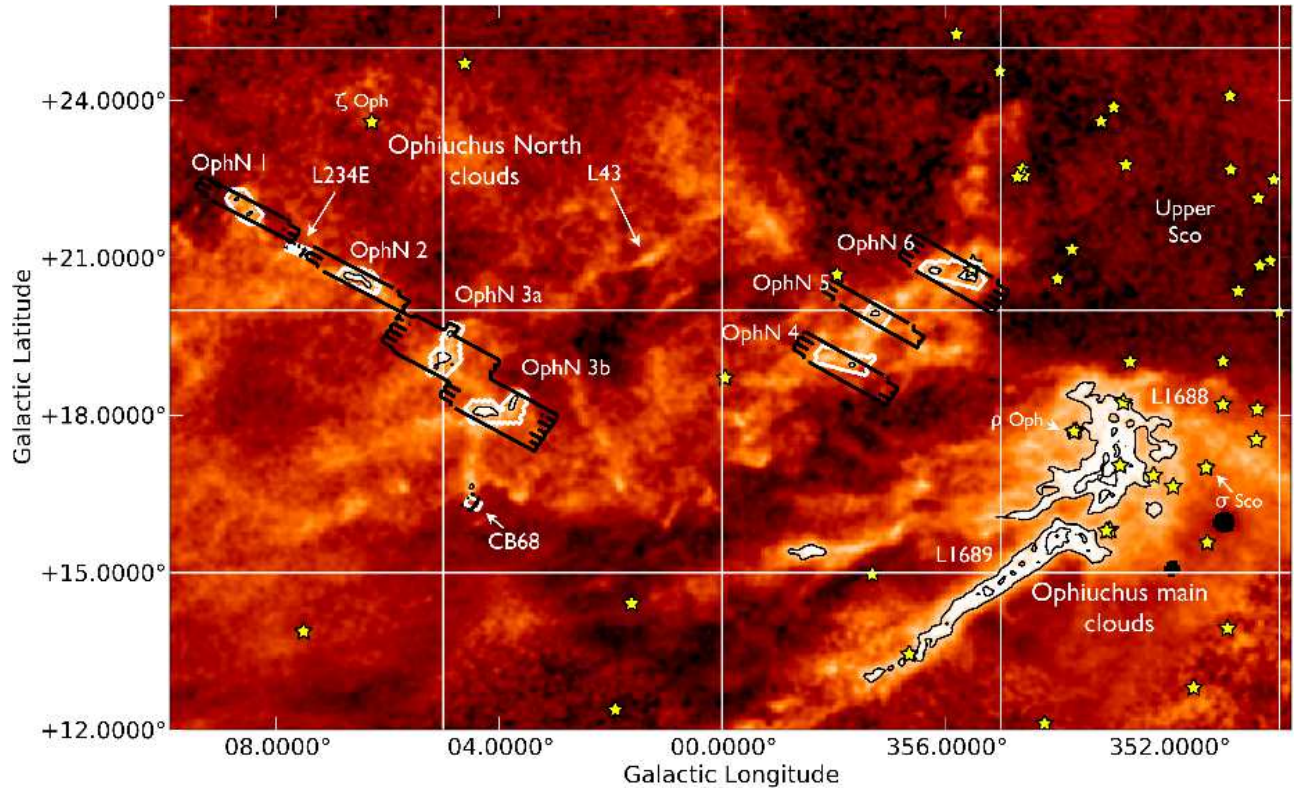}
\caption{Regions observed by IRAC (white boxes) and MIPS (black boxes) overlaid on visual extinction from \protect\citet{dobashi05}.  Extinction contours are at $A_V=3$ and 6. OB stars associated with Upper~Sco are marked by stars (\citealt{mamajek08}). \label{fig:coverage}}
 %Green circles mark the C$^{18}$O cores \protect\citep{tachihara00a}.  ADD  NOZAWA/TACHIHARA CORE LABELS?}
\end{figure}
\end{landscape}

The \SpitzerGB\ program aimed to complete the mapping of local star formation started by the \Spitzer\ ``From Molecular Cores to Planet-forming Disks'' (c2d) project \citep{c2d,evans09} by targetting the regions IC5146, CrA, Scorpius (renamed Ophiuchus~North), Lupus II/V/VI, Auriga, Cepheus Flare, Aquila (including Serpens South), Musca, and Chameleon to the same sensitivity and using the same reduction pipeline \citep{gutermuth08,harvey08,kirk09,peterson11,spezzi11}.   Images were made at 3.6/4.5/5.8/8.0\micron\ with the Infrared Array Camera (IRAC; \citealt{fazio04}) and 24,70 and 160\micron\ with the Multiband Imaging Photometer for \Spitzer\ (MIPS; \citealt{rieke04}).  With an 85cm mirror, IRAC observes with an angular resolution of $2''$ whereas MIPS is diffraction limited with $6''$, $18''$ and $40''$ resolution at 24, 70 and 160\micron\ respectively.  For our observations, we targetted small regions encompassing the $A_V>3$ contours from the optical extinction map of \citet{dobashi05}, as shown in Fig.~\ref{fig:coverage}.  The area in Oph~N which lies above $A_V>3$ is fragmentary and scattered over an area of 20 sq. deg.   Some of these regions (L158,L204,L146/CB68,L234E,L260,L43) had already been mapped as part of the \Spitzer\ ``Cores to Disks'' project \citep{c2d}.  These were avoided by the \Spitzer\ Gould Belt team to avoid unnecessary duplication of observations, as the two projects work to the same target sensitivities.  Most of these c2d data are incorporated in this study of Oph~N.  The exception is L43 which is presented separately by \cite{chen09}.  Ultimately, seven new areas were mapped by \SpitzerGB\ with IRAC and MIPS.  The \citet{dobashi05} and \citet{rowlesfroebrich09} extinction maps, which are not limited to the IRAC observations but extend across the entire area covered by MIPS, confirm that all $A_V > 3$ \citep[measuring from ][]{dobashi05} or $A_V > 4$ \citep{rowlesfroebrich09} regions in these filaments were observed by IRAC with the exception of two small clouds $0.3^\circ$ to the south of OphN~6 (included in the MIPS map) and $0.4^\circ$ to the north of CB68 \citep[][ core q2]{tachihara00a}.   Details of the datasets included for each region are listed in Table~\ref{tbl:aors}, including the observation dates, Astronomical Observation Request  (AOR) identification numbers, program identification (30574 for \Spitzer\ Gould Belt, 139 for ``Cores to Disks''), and duration.  Table~\ref{tbl:assc} gives the associated Lynds Dark Clouds \citep{lynds62} and molecular cores \citep[C$^{18}$O, ][]{tachihara00b} for each region.

The overlap area covered in all four IRAC bands is slightly smaller than the area covered in any single band because the array design leads to an offset between the 3.6/5.8\micron\ and 4.5/8.0\micron\ maps. The final area covered by MIPS is much larger than with IRAC because of the long MIPS scans, with almost all the IRAC areas covered by MIPS as shown in Fig.~\ref{fig:coverage}.  In total, 2.1 square degrees were covered by IRAC and 6.5 with MIPS.  This is roughly half the area of 14.4~sq.deg. covered by MIPS in the main Ophiuchus clouds \citep{padgett08}.

The observations of each area were split between two epochs to allow removal of transient objects, with the second epoch maps offset from the first epoch maps so that bad pixels do not always lie in the same sky positions.  The IRAC observations were taken with a total integration time of 48 seconds per point, split equally between the two epochs, and an offset of $10''$ in array coordinates between epochs.  Short integrations in High Dynamic Range (HDR) mode were also taken for all regions in which bright YSOs were expected (the exception was L234E).  The \SpitzerGB\ MIPS observations were taken in fast-scanning mode, stepping by $240''$ cross-scan to fill gaps in the coverage, and ($125''$,$80''$) between the two epochs in order to provide full 70\micron\ coverage with only half the array working.    MIPS total integration times were 32.4s at 24 and 70\micron\ and 6.2 seconds at 160\micron. % from Spot AORs 
The ``Cores to Disks'' observations of the small cores L234E and CB68 were taken in MIPS photometry mode.  Each core was observed in 2 epochs.  At 24\micron\, 1 cycle of 3 seconds was taken at each epoch for a total integration time of 84s.  At 70\micron\, 3 cycles of 3 seconds were taken at each epoch for a total integration time of $\sim 94$s.
% Table of associated dark clouds
% from Dobashi atlas

\subsection{Data reduction}

% brief description of pipeline (Evans et al. 2007, Rebull et al. 2007, Young et al. 2005)
% Image problems and corrections - now standard enough to refer back to earlier papers?
% Mosaicing and source extraction

Data reduction was carried out using the c2d pipeline as described in the delivery documentation (\citealt{c2ddel}; see also \citealt{harvey06} and \citealt{rebull07} for details of IRAC and MIPS processing).    The Basic Calibrated Data (BCD) from the \Spitzer\ Science Center were processed by the \SpitzerGB\ team to remove artifacts ( eg. bad pixels, jailbar effects, muxbleed column pulldown due to bright sources) and apply the location-dependent photometric corrections. 

\subsubsection{Mosaics}

Mosaicking was carried out using the SSC ``Mopex'' code.  For IRAC mosaics,  the high dynamic range (HDR) data were included in the final map, which improves the dynamic range and allows for the inclusion of otherwise saturated sources at the expense of slightly increased noise levels.  For both IRAC and MIPS, both epochs were combined.  

The \SpitzerGB\ Oph~N data were reduced to form six separate maps, numbered OphN~1--OphN~6 in order of decreasing Galactic longitude.  These were originally mapped as Sco~1--6 but relabelled `OphN' in line with the overall cloud nomenclature change.  OphN~1,4,5 and 6 each incorporated a single \SpitzerGB\ AOR area.  For OphN~2, the c2d cloud L204C was mosaicked with the \SpitzerGB\ data.   For OphN~3, the MIPS observations for two of the \SpitzerGB\ regions and the overlapping c2d region L158 were merged into a single mosaic.  The IRAC observations for OphN~3 remain split into OphN~3a (including L158) and OphN~3b.    The two separate c2d cores L234E and CB68 were mosaicked individually.  These pairings are indicated in Table~\ref{tbl:aors}.    A finding chart showing the six \SpitzerGB\ regions and the two complementary c2d areas overlaid on the \citet{dobashi05} extinction map is shown in Fig.~\ref{fig:coverage}.

\subsubsection{Catalogues}
\label{sect:catalogues}
Source extraction was carried out in each of the 4 IRAC bands on the combined mosaics (2 epoch plus HDR), plus MIPS 24\micron\ on the combined epoch mosaics, using c2dphot, a photometry tool based on DoPHOT \citep{harvey06, schechter93}.  Sources believed to be real at one or more wavelengths were band-filled at the missing wavelengths, first in IRAC bands 1--4 then MIPS~24\micron\ then MIPS~70\micron\ by fitting a point spread function (PSF) to images at the missing wavelengths, and these fluxes are included in the catalogues with bandfilling noted in the data quality flags.  Fluxes were extracted from the 2MASS point source catalogue \citep{skrutskie06} using a $2''$ position matching criterion.  The 70\micron\ source list was separately matched to the shorter wavelength catalogue with an $8''$ position matching criterion.  With a larger pixel size of $10\arcsec$
%$5.20''$ 
at 70\micron\ (compared to $1.2\arcsec$ for IRAC and $2.55\arcsec$ at 24\micron)  sources often matched more than one shorter-wavelength source.  The 70\micron\ fluxes were assigned to the best candidate by hand matching the spectral energy distribution (SED), with the remaining possible 70\micron\ emitters also noted in the catalogue.  Source matching was not attempted at 160\micron\ as the spatial resolution is low at this wavelength ($\sim 40''$)
%pixel dimensions are $16''\times18''$)
 and most bright regions are saturated.  For further details on source extraction see the c2d delivery documentation \citep{c2ddel}.

\subsubsection{160 micron data}

The 160\micron\ data obtained in fast scan mode mapping does not have enough redundancy to fill in completely all the gaps due to a dead readout and the intermediate gap between the array detector rows \citep{mipshandbook}. Furthermore with only 2 epochs, effectively 2 pointings per pixel, the effects of hard radiation hits and saturation translates into small regions without data. To deal with these gaps and to preserve as much as possible of the diffuse emission, the images are first resampled from 15\arcsec\ to 8\arcsec\ pixels and then a 5~pixel by 5~pixel median filter is applied to the image. The net effect is a slight redistribution of surface brightness ($\sim 15\%$) and smearing of the original beam from 40\arcsec\  to about 1\arcmin\ in size.   160\micron\ data are not available for the two c2d cores CB68 and L234E.  

%(Mips Instrument Handbook section 6.5 there are 5 dead pixels at 160um, need 3 or 4 epochs to completely recover the %missing pixels.)

Fluxes at 160\micron\ are determined using aperture photometry with a 32\arcsec\ radius aperture, an annulus from 64-128\arcsec\, and an aperture correction of 1.97.  Flux density uncertainties are about 20\% below 5~Jy and 30\% for higher flux densities \citep{rebull10}. Visual inspection of each 160\micron\ source in the original unsmoothed image is used to determine whether the object is cleanly detected or contaminated by data dropouts.

% (IF NEEDED The zero-point for MIPS-160 is 0.159 Jy, from the SSC website.)

%\subsection{Data quality}

% Image quality 
% Completeness and reliability
%            Completeness is estimated by Monte-Carlo analysis with artificial stars

%\subsection{High-reliability bandmerged catalogue}

% Reliability criteria - "most reliable" subcatalogue trimmed to 7 sigma, removed extended sources, saturated sources.

\section{Results}
\label{sect:results}

% Figures:
% Area coverage on eg. extinction greyscale
% Mosaics in each band
% IRAC 3-colour images (optionally with extinction contours), zooms of interesting bits.
% Colour-colour/colour-magnitude diagrams.
% IRAC+MIPS 3-colour plots

% TABLES
% Number of sources extracted 
% Band-merged sourcelists.
% Detection statistics

\begin{landscape}
\begin{figure}
\epsscale{0.9}
%\plotone{PS/ophn_rgb_124.eps}
\plotone{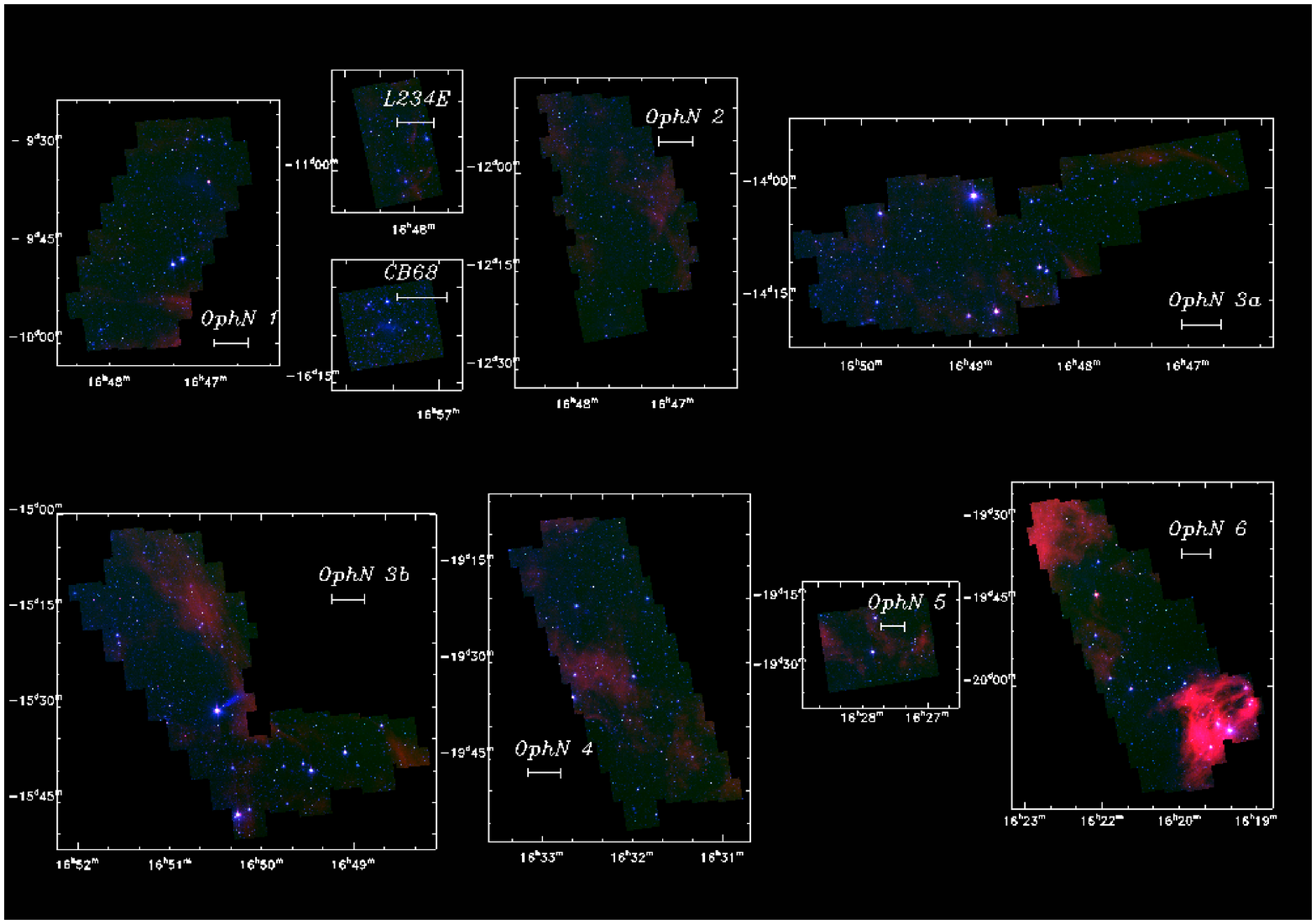}
\caption{Three-colour mosaics of the Oph~N cloud regions imaged by \Spitzer\ with IRAC 3.6\micron\ (blue), IRAC~4.5\micron\ (green) and IRAC~8.0\micron\ (red).  Only the overlap regions imaged at all three wavelengths are shown.  The scale bars represent 0.2~pc at the assumed distance of 130~pc. \label{fig:rgbirac}}
\end{figure}
\end{landscape}

\begin{landscape}
\begin{figure}
%\begin{tabular}{c}
%\includegraphics[scale=0.7,angle=0]{PS/sco_rgb_124.eps}\\
%\includegraphics[scale=0.7,angle=0]{PS/sco_rgb_245_crop.eps}
%\includegraphics[scale=0.6]{PS/sco_rgb_124.eps}\\
%\includegraphics[scale=0.9]{PS/ophn_rgb_245.eps} % original EPS from IDL.  Colour.
%\includegraphics[scale=0.9]{f3.eps} % original EPS from IDL.  Colour.
\epsscale{0.9}
\plotone{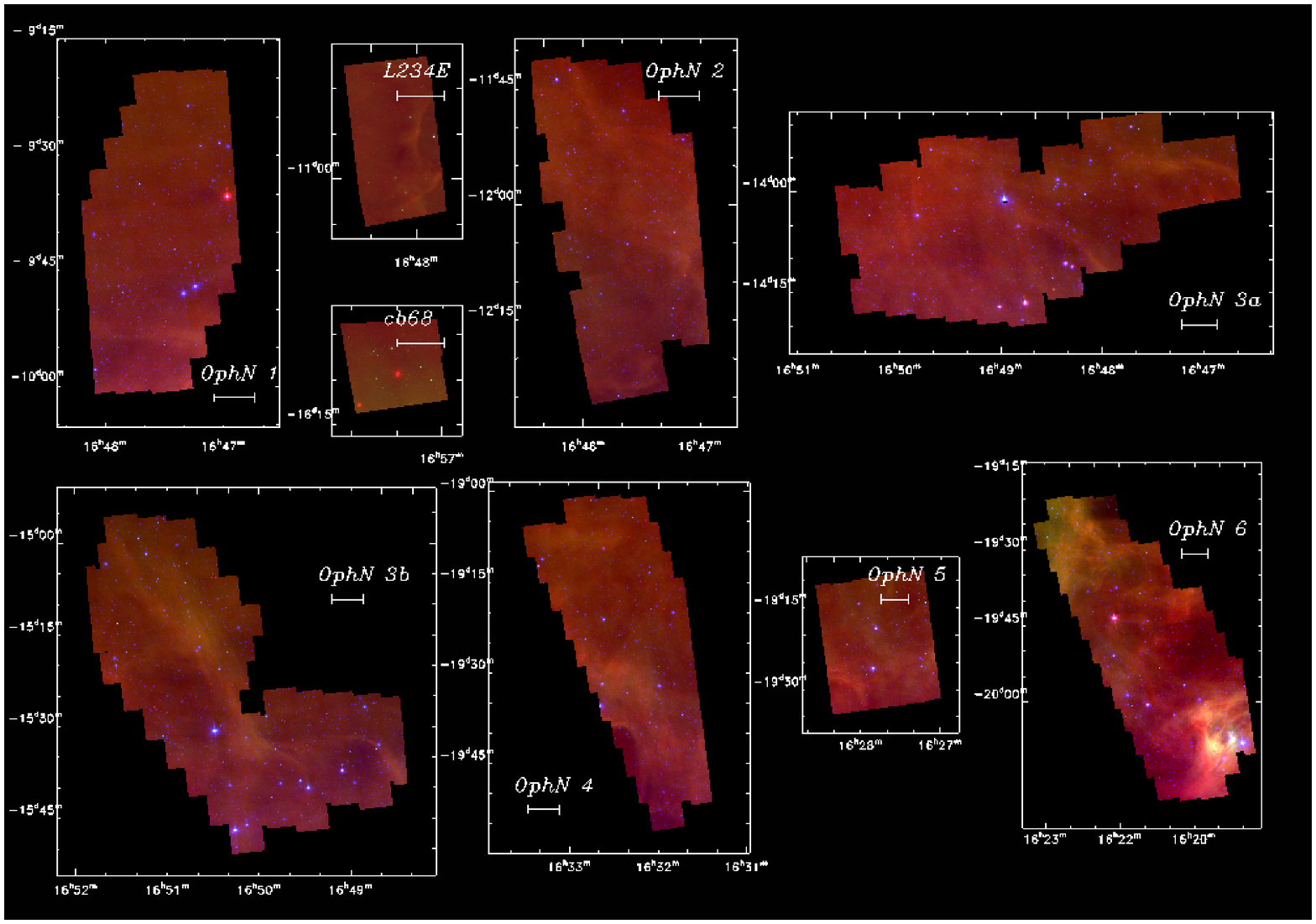}
%\end{tabular}
 \caption{Three-colour mosaics of the Oph~N cloud regions imaged by \Spitzer\ with IRAC 4.5\micron\ (blue), IRAC~8.0\micron\ (green) and MIPS~24\micron\ (red).   Only the overlap regions imaged at all three wavelengths are shown.  The scale bars represent 0.2~pc at the assumed distance of 130~pc. \label{fig:rgbmips}}

\end{figure}
\end{landscape}

Red-green-blue (RGB) images for the eight Ophiuchus~North regions mapped (OphN~1--6, CB68 and L234E) are shown in Figs.~\ref{fig:rgbirac} and \ref{fig:rgbmips}.  The first figure shows a combination of IRAC 3.6\micron\ (blue), 4.5\micron\ (green) and 8\micron\ (red), and the second combines IRAC~4.5\micron\ (blue), IRAC~8\micron\ (green) and MIPS~24\micron\ (red).  The RGB images each cover the overlap region where data is available from all three chosen \Spitzer\ instruments, corresponding to the IRAC areas (white boxes) in Fig.~\ref{fig:coverage}.  The regions are irregular with sizes starting at 0.3~pc for CB68 and L234E, with OphN~3a the largest region mapped at a couple of parsecs in length; scale bars of $0.2$~pc are shown on Figs.~\ref{fig:rgbirac} and \ref{fig:rgbmips}.

\jh{Unextincted foreground stars appear blue-white in these images.  Embedded protostars have redder colours, as do background galaxies, background giants, and planetary nebulae, extincted main-sequence stars and foreground late-type stars \citep{robitaille08}.} The red wisps of 8\micron\ emission are PAH emission from heated clouds.  Shocked H$_2$ emission from molecular outflows can show up at 4.5\micron\ in green on the IRAC 3-colour plot, but there is no evidence for this here. 

Figs.~\ref{fig:rgbirac} and \ref{fig:rgbmips} shows that there are no dense protostellar clusters in Ophuchus~North, such as \Spitzer\ found in Serpens and Auriga \citep{gutermuth08,matthews12}.  The bright region in the southwest of OphN~6 is the optical reflection nebula IC4601 \citep{magakian03}, excited by a few late-type B and A stars (see Sect.~\ref{sect:sco6}).  There is no dense young reddened cluster associated with these intermediate-mass stars, as Fig.~\ref{fig:rgbmips} shows. The young stellar objects associated with OphN~6 and other regions are discussed in Sect.~\ref{sect:ysos}.   

\subsection{Source statistics}
\label{sect:stats}

% Background contamination (comparison with SWIRE sources in colour-colour plots)
% Colour-colour and colour-magnitude plots using IRAC and 2MASS bands inc. SWIRE sources, reddening vectors (Huard et al. ??).  AGB stars
% MIPS: 24 micron source counts
% MIPS: colour-magnitude diagrams

The Ophiuchus~North IRAC and 2MASS source detection statistics are given in Table~\ref{tbl:detections}.  In the area covered by all four IRAC bands, just over 100,000 sources were detected in at least one IRAC band with a signal-to-noise of at least 3, with more than 5000 detected in all four IRAC bands.   Nearly all of the 2MASS sources in the field were also detected by IRAC, with half of these detected in all 4 IRAC bands.  The number of sources detected as a function of the required signal-to-noise in the individual wavelength maps is given in Table~\ref{tbl:sn}.  These counts include the entire region covered in each waveband and not solely the overlap area, which is why the 3.6\micron\ detection count for S/N=3 is larger than in Table~\ref{tbl:detections}.

\subsection{Background sources}
\label{sect:background}

% Spectral index alpha (from IRAC + 2MASS) and numbers of Class I,flat,II,III sources.
% MIPS-only classification: Ks-[24] Young05, Rebull07,Chapman07. (eg. for regions outside IRAC) - TBD
% Fuzzy 2MASS+IRAC+MIPS classification (Harvey et al. Serpens synthesis).

\begin{figure}
\epsscale{0.5}
%\plotone{PS/sco_diff.eps} % original EPS from IDL, BW only.
\plotone{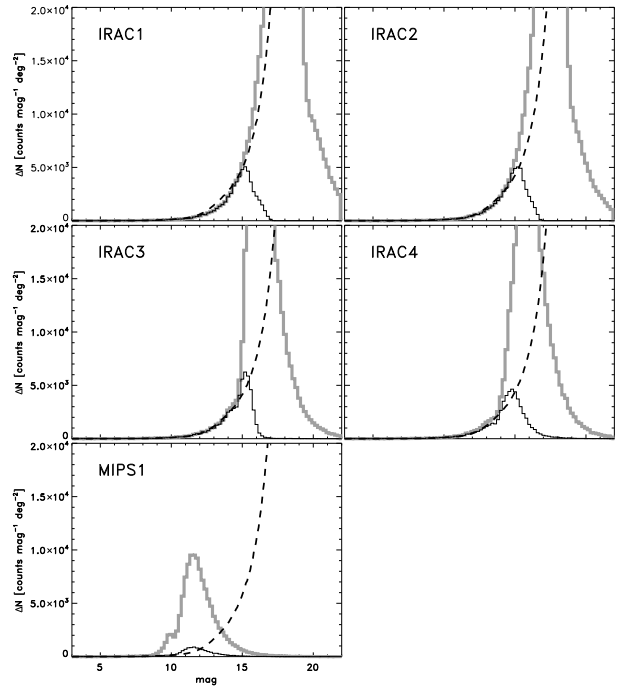} % original EPS from IDL, BW only.
 \caption{\jh{ Differential source counts compared to the Wainscoat model for the galactic stellar background.  The thick grey line shows the differential source counts per square degree in each band for all catalogued sources.  The thin black line shows the differential source count for only the sources classified as stars.  The stellar count can be compared to the dashed line which shows the Wainscoat model for the stellar density in the Galaxy.} \label{fig:wainscoat}}

\end{figure}

Most of the sources detected by \Spitzer\ are expected to be either main-sequence stars or background galaxies.  In order to estimate the fraction of young sources associated locally with the Oph~N star-forming region, a comparison with the Wainscoat model for the Galactic stellar background at this latitude is made in Fig.~\ref{fig:wainscoat} \citep{wainscoat92}.  The figure shows differential source counts for all sources in each of the 4 IRAC bands and MIPS1 from the Oph~N catalogues (grey curves for all sources; black curves for \jh{sources classified as stars} with $3\sigma$ detections) compared to the Wainscoat predictions (dashed curve).  For Oph~N, most of the source detections are main-sequence stars, as shown by the lack of excess counts over the Wainscoat model in the IRAC bands.  The excess in the MIPS band over the Wainscoat model is mainly due to background galaxies.  The completeness limit at approximately 15~magnitudes in the IRAC bands can be seen in the turnover in the star counts (black line).

\subsection{YSO candidates}
\label{sect:ysos}

\begin{figure}
\epsscale{0.8}
%\plotone{PS/swire_comparison_plots.eps} % original ps from IDL.  Colour required.
\plotone{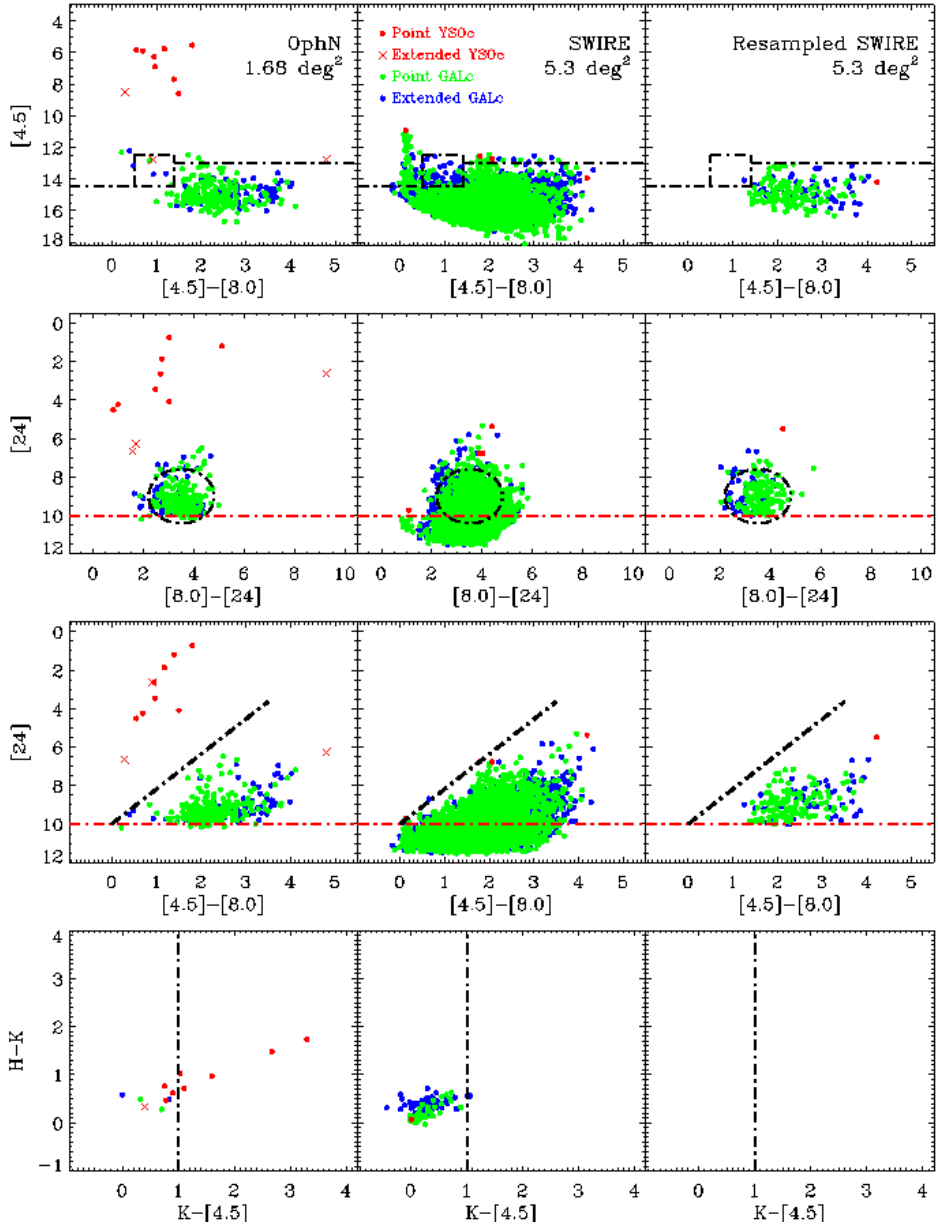} % original ps from IDL.  Colour required.
\caption{Colour-magnitude diagrams for all Oph~N regions combined, showing the YSOc identification criteria from \protect\citet{harvey07} and \protect\citet{c2ddel}.  Symbols are pointlike/extended YSOcS (red circles/crosses) and pointlike/extended galaxies (green/purple) as shown in the key.  The black lines show the fuzzy colour-magnitude cuts that define the YSO candidate criterion.  Objects fainter than the hard limits shown as red dot-dashed lines are excluded from the YSOc category.  The area covered is the $1.68 \hbox{deg}^2$ covered by both IRAC and MIPS.\label{fig:ysosel}}
\end{figure}

The main aim of the \SpitzerGB\ programme is to locate the young stellar objects associated with star formation.   In the Oph~N molecular clouds, eleven candidate Young Stellar Objects (YSOc) were identified on the basis of their position in IRAC colour-magnitude diagrams, as shown for each of the IRAC 4-band detections in Fig.~\ref{fig:ysosel}.  The selection process is described in detail in \citet{harvey07} but in summary, each colour-magnitude diagram yields a probability of a source being a YSOc with the region occupied by galaxies, based on the SWIRE sample, giving a low probability. The individual probabilities are then combined with some additional criteria to separate the candidate YSOs (coloured red) from galaxies.   The method aims to eliminate background sources at the expense of also eliminating some YSOcs.  This process is not infallible and followup visual inspection of the mosaics identified one extended and elliptical YSO candidate as a galaxy (6dFGS gJ164828.8 -141437; \citealt{6dFGS}).

The remaining ten YSOc were classified Class~I, flat-spectrum, II, or III according to their spectral index fit between K-band (1.2\micron) and MIPS band 1 (24\micron) as listed in Table~\ref{tbl:ysos}.  Fluxes for these sources are listed in Table~\ref{tbl:fluxes}.  Spectral energy distributions (SEDs) for these sources are shown in Figure~\ref{fig:seds}.   \jh{These include optical ($B$,$V$,$R_\mathrm{C}$) magnitudes from the NOMAD catalogue from a $5''$ radius position search \citep{nomad}.}  For the more evolved sources, the SEDs are also shown dereddened to fit a stellar photosphere at the short wavelengths.  \jh{The best wavelength/band for identifying the continuum of stellar emission in these young objects is $J$-band, which was used to normalize the reddened photospheres to the SEDs.}  In all but two cases this was a K7 star, but for sources 3 and 4 an A0 star was needed in order to reproduce the high fluxes.  These YSOc SEDs show disk excesses over a stellar photosphere at long wavelengths.  \jh{The poor match between the SED and the optical data in some cases (YSOc~3,6,7) is most likely to be due to misassociations as several optical sources within the $5''$ radius contribute to the SED of a single source in the Spitzer bands.   Some of these sources could also have ultraviolet excesses due to disk accretion that prevent good fits to main sequence stellar photosphere models \citep{calvet98}.  It is also possible that the reddening law in these dense cores could differ from the $R_V = 5.5$ Weingartner \& Draine interstellar grain model \citep{weingartnerdraine01, indebetouw05}, but this correction would be likely to be similar for all sources. }

No photosphere is fitted to the flat-spectrum or Class~I sources.  These embedded YSOcs have emission peaking at longer wavelengths than main-sequence stars and can be identified by their redder colours in the RGB diagrams (Figs.~\ref{fig:rgbirac} and \ref{fig:rgbmips}).  The Class~I sources in OphN~1 (YSOc~1) and CB68 (YSOc~10), and the flat-spectrum source in OphN~6 (YSOc~8), show up in this way.  

More than half of the YSOcs were detected by IRAS and all but three were previously identified as YSOs (see Table~\ref{tbl:ysos}).  The two deeply embedded Class~0/I sources are the known protostars L260~SMM1\citep{am94,visser02} and CB68\citep{carballo92,huard99}.  Most of the Class~II objects are T~Tauri stars which have been the subject of multiple studies.  The three new identifications are the Class~III sources OphN~YSO6, 7 and 9, identified through their small 24\micron\ excesses.   Two of these (YSOc6 and 7) may be background AGB stars; we examine this possibility in Sect.~\ref{sect:agb}.  OphN~7 and 9 both lie between the extinction peaks of OphN~6 (Fig.~\ref{fig:extinction}).  Finding pre-main-sequence stars in OphN~6 is unsurprising as it lies on the edge of the Sco-Cen association.

\subsection{24\micron\ emission and MIPS-only YSO candidates}

A larger area of Oph~N was covered by MIPS at 24, 70 and 160\micron\ and by 2MASS in the near infrared, but not at intermediate wavelengths by IRAC.  YSOs in that region are not found on the standard IRAC-based colour-magnitude diagrams as shown in Fig.~\ref{fig:ysosel}.  A start on identifying YSO candidates with no IRAC data can be made by looking for sources with red colours in 2MASS $K_S-[24]$.

The $K_S$ vs. $K_S-[24]$ colour-magnitude diagram is shown in Fig.~\ref{fig:kmips}, following \citet{rebull07} (Perseus) and \citet{padgett08} (Ophiuchus).  All Oph~N sources detected at these wavelengths are included (OphN~1--6, CB68 and L234E combined).  The source symbols are based on the c2d pipeline classification.  To be classified as a YSOc requires fluxes in all four IRAC bands and MIPS1; red sources which do not fulfil this criterion (eg. those not observed by IRAC) tend to be classified as `galc' or `star+dust' (\citealt{c2ddel}; see Sect.~\ref{sect:catalogues}).

In Fig.~\ref{fig:kmips}, sources with the colours of stellar photospheres appear to the left with $K_S-[24]<2$, and most of the sources in this region are already classified as stars (grey circles) by SED fitting.   A reddening vector is shown in the bottom left of the plot, assuming an extinction law of $A_{24}/A_{K_S} = 0.5$ as recommended by \citet{chapman09}; however, the 24\micron\ relative extinction is quite uncertain.  Reddened stars are likely to appear down and to the right of the main stellar population.  The extinction in Oph~N is low over most of the cloud, with only a small fraction above $A_{\mathrm v} = 3$ (Figs.~\ref{fig:coverage} and Sect.~\ref{sect:extinction}), so few  background stars should be reddened significantly, and indeed there are relatively few sources in this region.  The lower left region of the diagram is not populated because stellar photospheres with $K_S>10$ do not lie above the MIPS 24\micron\ sensitivity limit unless they are reddened.

Background galaxies (`Galc', green $\times$) typically have low $K_S$ fluxes and lie towards the bottom of the plot, as shown by the colours of objects in the SWIRE catalogue \citep{rebull07,padgett08}.  In this region are found most of the c2d pipeline-classified `star+dust' sources (red $+$), which have the colours of a stellar photosphere but excess due to dust at the long wavelengths.  

Candidate YSOs with $K_S-[24]$ excesses appear in the centre of the plot, to the right of the stars, depending on the contribution of dust in the disk and/or envelope.  This is where the ten `YSOc' already identified by the c2d pipeline and IRAC colour-magnitude diagrams appear (red filled circles), with the exception of CB68, which was not detected by 2MASS (or IRAC1) and does not appear on this plot.  In this region there are several additional YSO candidates with no IRAC detection, currently classified `star+dust' (red crosses) due to their long-wavelength excess over a stellar photosphere.  

At this point we make a selection for a clean sample of MIPS-only YSO candidates, rejecting the regions of colour-magnitude space with $K_S<13$, which excludes the majority of background galaxies, $K_S-[24]>2$, which excludes the background stars, and $[24]>9$ which removes reddened stars.  This is a conservative selection which probably rejects some genuine YSOs.  We make the galaxy cut at $K_S=13$ rather than $K_S=14$ \citep{rebull07,padgett08} because the two brightest sources below $K_S=13$ in the flat-spectrum region of the plot both have galaxy identifications.  The misclassified `YSOc', which has an IRAC detection and appears as a red filled circle, is 6dFGS~gJ164828.8$-$141437 and the brighter $K_S\simeq 13$ source is 6dFGS~gJ163139.8$-$202044.    

The five MIPS-only YSO candidates in Oph~N selected by these criteria are listed in Table~\ref{tbl:mipsysocs} .  We can classify them further based on the standard spectral index $\alpha$ calculated from $K_S-[24]$ into Class~I, flat-spectrum, Class~II and Class~III sources \citep{rebull07,greene94}; the corresponding regions of colour space are identified on Figure~\ref{fig:kmips}.    The three Class~II candidates M5, M2 and M4, in OphN~3,4 and 6 respectively, are close in colour to the existing YSOcs and have strong $K_S$ and 70\micron\ detections.  The fainter $K_S$ source M3 is also a Class~II candidate in OphN~6.  In OphN~3, the Class~III source M1 which lies redwards of the main stellar population is identified as IRAS~16484$-$1557.  

To fill in the missing mid-IR portion of the spectrum for regions not covered by IRAC, additional mid-infrared fluxes were taken from the WISE Preliminary Release Source Catalog \citep{WISE}\footnote{http://wise2.ipac.caltech.edu/docs/release/prelim/expsup/}.  WISE provides fluxes at 3.4, 4.6, 12 and 22\micron\ with angular resolution of 6--$12''$.  The resulting spectral energy distributions, shown in Fig.~\ref{fig:kmipsseds}, confirm the sources to be similar to the Class~II/III YSOcs in Fig.~\ref{fig:seds}.   The classification of the source M5 (V1121~Oph), which shows a strong disk excess, deserves some further comment.  Aperture photometry with a $100''$ aperture gives a 24\micron\ flux an order of magnitude higher than the c2d pipeline ($\sim 4000$ rather $\sim 400$~mJy) and the higher flux is confirmed by WISE at 22\micron\ ($5915\pm49$~mJy); see Table.~\ref{tbl:mipsfluxes}).  The higher 22/24\micron\ fluxes move this source to the Class~II/flat-spectrum boundary (Fig.~\ref{fig:kmips}).

$K_S-[24]$ alone is an imprecise selection tool, and there may be additional YSOs in the excluded regions, particularly among the faint 24\micron\ sources.  Colour-magnitude searches based on WISE data may be a way to identify these, but are beyond the scope of this paper.

\subsection{AGB star contamination}
\label{sect:agb}

There remains the possibility that some of the YSOcs may be background asymptotic giant branch (AGB) stars.  Like YSOs, AGB stars have red colours due to the dust shells which surround them.  They occupy a similar region in colour-magnitude space to YSOcs, and cannot easily be separated by the criteria shown in  Fig.~\ref{fig:ysosel} \citep{robitaille08}.  As befits an old stellar population, AGB stars are distributed throughout the Galaxy with increased counts in the thick disk and bulge.  Oph~N lies 17--$24\deg$ from the plane, so using the Galactic distribution model from \citet{ishihara11} an average AGB contamination of less than 0.5~deg$^{-2}$ would be expected, or fewer than $1$ AGB star in the 2.1~deg$^{2}$ area covered by IRAC and fewer than $3$ \jh{AGB stars} in the 6.5~deg$^2$ regions covered by MIPS.

AGB stars can be identified by their low proper motions,  reflecting their Galactic distances and inconsistent with membership of nearby Upper~Sco.  From Hipparcos, Upper~Sco has average proper motions of  $(\mu_{\alpha}, \mu_{\delta}) = (-11, -24)$ in Galactic coordinates \citep{dezeeuw99,mamajek08}.  
%$\mu_{\alpha\cos\delta}, \mu_{\delta} ~%= -11, -24$ in Galactic coordinates \citep{dezeeuw99,mamajek08}.  
Two of our IRAC+MIPS1 selected YSOc have inconsistent and low proper motions:  YSOc~6 (IRAS~16191$-$1936 in OphN~3b) has proper motions of $4.7 \pm 7.2, -1.2 \pm 7.2$~mas/yr \citep{PPMXL} and YSOc~7 (OphN~6) has proper motions of $-2.6\pm5.5, 2.3\pm5.5$~mas/yr.  Neither of these are consistent with a nearby location in Oph~N, whereas the other YSOcs all show proper motions consistent with Upper~Sco.  

Additionally, AGB show a steep spectral index at long wavelengths: a criterion of $[8.0]-[24] < 2.2$ can be used to select them, though this criterion alone will not distinguish them from YSOcs \citep{whitney08,robitaille08}.   Fig.~\ref{fig:agbcol} shows several combinations of colours and magnitudes of the YSO and AGB candidates compared to those of the known AGB sample in Serpens (identified by infrared spectroscopy; \citealp{oliveira09}).  It can be seen that the two IRAC-observed AGB candidates in Oph~N lie in the same region in colour-magnitude space as the confirmed AGB stars.  YSOc~6 (IRAS~16191$-$1936 in OphN~3b) has $[8.0]-[24] = 1.2$ and YSOc~7 (OphN~6) has $[8.0] - [24] = 1.0$.   These two sources are among the most luminous in our sample, as shown in Fig.~\ref{fig:seds}.   
In Fig.~\ref{fig:agbcol}, YSOc~9 also has the colours of an AGB candidate, but its proper motions are consistent with membership of Upper~Sco ($\mu_{\alpha}, \mu_{\delta} = -3.9\pm4.9, -29.9\pm4.9$~mas/yr; \citealp{PPMXL}).

In addition, two of the MIPS-selected YSO candidates are likely to be background AGB stars.  Sources M1 and M2 both have low proper motions inconsistent with Upper~Sco: M1 ($ \mu_{\alpha}, \mu_{\delta} = -4.7\pm4.9, -1.0\pm4.9$~mas/yr) and M2 ($-4.7\pm4.9, -1.0\pm4.9$~mas/yr; \citealp{PPMXL}).  IRAC fluxes at 8\micron\ are not available to apply the Whitney et al.~$[8.0]-[24] < 2.2$ colour criterion exactly, but applying the same principle to WISE fluxes these two sources have the steepest spectra of the sample between 12 and 22\micron\ ($[12]-[22] < 1.0$), supporting their identification as AGB stars.  In the ASAS variable stars catalogue, M1 is identified as a Mira variable with a magnitude variation of 1.9~mag and period 274.050380 days (ASAS 165122-1602.9, \citealp{pojmanski05}).

The four AGB candidates among the YSOcs are identified by footnotes in Tables~\ref{tbl:ysos} and \ref{tbl:mipsysocs} and in Figs~\ref{fig:scuba}--\ref{fig:excess}.  They are not counted in our final YSOc tally (Sect.~\ref{sect:discussion}).

\subsection{70 micron emission and SCUBA maps}

Nine out of the ten YSOcs listed in Table~\ref{tbl:ysos}  and two out of four of the MIPS-only YSOcs in Table~\ref{tbl:mipsysocs} were observed and detected at 70\micron.  Detections at 70\micron\ are either a sign of disk excess or a dusty envelope, as is clear from the SEDs (Fig.~\ref{fig:seds}).  Often, these sources can also be detected in the millimetre/submillimetre, on the long-wavelength side of the peak of the SED.  In Fig.~\ref{fig:scuba}, we show the MIPS 70\micron\ maps overlaid with contours of submillimetre emission where available.  The 70\micron\ maps show striping artifacts remaining from the reconstruction \citep{c2ddel}.  We searched the re-reduced SCUBA archive for 850\micron\ maps associated with the Oph~N regions \citep{difrancesco08} and found existing small SCUBA maps associated with parts of OphN~1 and OphN~3 (originally published in \citealt{visser02}), L234E \citep{kirk05}, and CB68 \citep{huard99,vallee03,young06}.  The two YSOcs in the south of OphN~3 and in OphN~6 do not appear to have been mapped by SCUBA, and we are not aware of any other mm/submm continuum observations of these sources.  

Of the two Class~I YSOcs, CB68 has a 70\micron\ detection: the OphN~1 Class~I lies just off the edge of the 70\micron\ map.  Both have been mapped and detected at 850\micron\ by SCUBA \citep{visser02,huard99}.  The flat-spectrum source (YSOc~8 in OphN~6) is also detected at 70\micron.  All but one of the Class~IIs are detected: the exception is the MIPS-only Class~II YSOc~M4 in OphN~6.  Neither the Class~III source YSOc9 nor the AGB candidates YSOc7 and M1 are detected at 70\micron, as expected from the mid-infrared falloff of the SEDs (Fig.~\ref{fig:seds} and \ref{fig:kmipsseds}).

% With IRAS and ISO: did SPITZER detect these objects (eg. Serpens)?
% With known YSOs (eg. Lupus)

\subsection{Extinction maps}
\label{sect:extinction}

Extinction maps provide an alternative measure of the cloud structure in Oph~N.   The \Spitzer\ Gould Belt catalogue includes a measure of the visual extinction $A_{\mathrm v}$ towards every source with a SED fitted by a reddened stellar photosphere (i.e. classified as a star).   By convolving the extinctions at each position with a Gaussian beam,  these values have been turned into extinction maps for all the area in Ophiuchus~North covered by IRAC.  The detailed procedure is explained in the c2d delivery document \citep{c2ddel}, but in summary, extinction maps are provided at a range of spatial resolutions, with the maximum resolution limited by the number of stars which contribute extinction measurements within each beam.  As this number decreases, the uncertainties in the extinction increase.  For the Gould Belt extinction maps to be accepted, a maximum of 10\% of beams at any given $A_{\mathrm v}$ level were allowed to be undersampled, defined as containing no stars within the beam FWHM.  Thus the lowest resolution Oph~N maps were sampled with a $300\arcsec$ beam but the highest resolution $A_{\mathrm v}$ maps vary from $90\arcsec$ (OphN~4 and 5) through $120\arcsec$ (OphN~1,3 and 6) to $150\arcsec$ in the high-extinction OphN~2.  The extinction maps are made primarily using stars with IRAC detections and at least one good 2MASS detection.  In addition, a limited number of stars with no 2MASS detection but good IRAC detections are used to supplement the measurements in the higher extinction regions ($A_V > 8.5$), where the number of near-infrared detections is reduced.

Extinction maps at $150''$ resolution (the highest resolution available for Oph~N regions 1--6) and $120''$ resolution (L234E, CB68) are plotted as black contours over the MIPS 24\micron\ mosaics in Fig.~\ref{fig:extinction}.  The advantage of the \Spitzer-derived extinction maps over the optical \citep{dobashi05} or near-infrared \citep{rowlesfroebrich09} is that they can probe high extinctions at relatively high resolution and so can detect compact, high column density cores.  This is particularly useful in Oph~N where extensive mm/submm maps are not yet available.
%, though this region forms part of the JCMT Gould Belt survey planned for SCUBA-2 \citep{gbs}.  

The mid-infrared-derived extinctions shown in Fig.~\ref{fig:extinction} are higher than the extinctions derived from 2MASS colours \citep{rowlesfroebrich09,froebrich05}  by typically 2 magnitudes and from optical star-count extinctions of \citet{dobashi05} by typically 3 magnitudes. This can be seen from the contours on the extinction maps in Fig.~\ref{fig:extinction}: the $A_V > 3$ regions from the \citet{dobashi05} star-count extinction maps (blue dashed contours) correspond roughly to $A_V > 4$ on the \citet{rowlesfroebrich09} near-IR extinction maps (shown as red dashed contours) or $A_V > 6$ on the \Spitzer\ extinction map (base level of black contours).    The IRAC selection (black boxes) was based on the \citet{dobashi05} $A_V = 3$ contour.   The reason for this difference in extinction level is not simple and requires further explanation.  A significant difference between the extinction maps produced for the \Spitzer\ Gould Belt survey compared to those for c2d is that no extinction offset is subtracted from the \Spitzer\ Gould Belt regions.  For the c2d maps, an extinction offset was calculated using off-cloud fields believed to be free of extinction from the molecular cloud.  These offsets, which for the Ophiuchus cloud lying nearest to Oph~N is $A_V=1.96$, were subtracted from the on-cloud extinctions.  It was believed that these corrections were for foreground, line-of sight extinction from the diffuse interstellar medium.  From a detailed comparison of the mid-infrared extinction law with the models \citep{chapman09} it is now thought instead that these extinction offsets are due to deviations from the assumed \citet{weingartnerdraine01} $R_V  = 5.5$ extinction law in the mid-infrared region of the spectrum when $A_V$ is low.  $A_V$ was calculated assuming the extinction law can be parameterised as $A_\lambda(R_V , \lambda)$ with $R_V = 5.5$ \citep{weingartnerdraine01,cardelli89}.  The conversion of mid-IR extinction to $A_V$ is nonlinear with the $R_V =5.5$ model producing reasonably reliable estimates of $A_V$ at high extinctions ($A_V >~ 10$) but overestimates at low extinctions ($A_V <~ 10$) by a factor which varies with $A_V$ and cloud.  This is the cause of the apparent extinction in the off-cloud fields, and the difference of $A_V= 2$--3 from the extinction maps derived at shorter wavelengths. 

% rf maps have resolutions 3.5-4' rising to 6' in the cores.  Using the median these maps are biased towards the lower extinctions in each area (A_K ~ A_V/100), and not sensitive to small high extinction regions.  RF maps show only two extinctions above Av~6, in Sco 3 and Sco 6.  Compare \Spitzer\ 150" = 2.5' resolution.

The high extinction regions are small and fragmented, typically less than a parsec in size.  The masses in each region above extinction thresholds of $A_V = 6, 10$ and $20$, derived from these maps, are given in Table~\ref{tbl:extinctionmasses}.  Approximately 100 solar masses of gas lies at $A_V > 10$ in Oph~N, split between 12 clumps (Fig.~\ref{fig:extinction}).  This is about 1/30 of the mass in Ophiuchus though similar to Lupus~I or IV \citep{heiderman10}.  Each individual clump only has a mass reservoir of order 10\Msun\,  sufficient to form a few low-mass stars,.  Several Oph~N regions show cores of $A_V\ge 20$  in the $150\arcsec$-resolution \Spitzer-derived extinction maps: there is one in OphN~1, two in OphN~2, and three in OphN~3 (Fig.~\ref{fig:extinction}).  The OphN~1 and CB68 cores contain the two Class~I YSOs OphN~YSO1 and 10, respectively, but the other extinction peaks are currently starless.  Class~II and III YSOs lie on the edges of the extinction regions in OphN~3a, 3b and OphN~6.

\subsection{160\micron\ and IRAS maps}
\label{sect:long}

Emission at wavelengths of 100\micron\ and longer is dominated by cold dust and extended cloud structure.  In addition to the MIPS 160\micron\ data, we also obtained IRAS 100~\micron\ images that were resolution enhanced using the HIRES algorithm\footnote{\tt http://irsa.ipac.caltech.edu/IRASdocs/hires\_over.html}. Image coordinates were additionally precessed from B1950 to J2000 coordinates using the Goddard IDL astron library. 
The algorithm improves the spatial resolution from the native IRAS ($3.8\arcmin \times 5.4\arcmin$) to $2\arcmin$ at 100~\micron\ wavelength \citep{cao97}. \jh{The MIPS 160\micron\ data have a factor 5 better spatial resolution of $40''$.} The HIRES processing provides sharper and more clearly defined filaments and other extended structure. Point sources are however surrounded by a ringing artifact. Since our focus is on the extended structure the point sources are removed from the 100~\micron\ image using median pixel replacement. 

%For alberto
%Fastscan mode, 240arsec cross-scan step, 2 epochs
%Ask Alberto how this compares with Taurus, or what the typical coverage is.
The \Spitzer\ 160\micron\ and IRAS~100\micron\ maps are shown side-by-side in Fig.~\ref{fig:long}.  The IRAS 100\micron\ maps have broader spatial coverage and show the continuous filamentary structure linking the OphN~1 to 3 clouds.  The OphN~4--5 clouds are also linked by 100\micron\ emission which takes the form of several cavities or bubbles.  Although there are ionising stars in the region, these lie to the north (the B2V star $\chi$~Oph at $l,b= 357.93,+20.68$ and, in the past Myr, $\zeta$~Oph; see Fig.~\ref{fig:coverage} of this paper and \citealp{nozawa91} Fig.~5) whereas the openings of the bubbles lie to the south.  The bubbles themselves contain no ionising stars to produce H{\sc II} regions.  A possibility is that they could be fluid instabilities at a hot gas / cold cloud interface where the flow from the Upper~Sco H{\sc I} bubble has broken out \citep[see discussion in][]{degeus92}.

The extended emission exhibits structure that is very similar at both 100 and 160 ~\micron, apart from resolution effects. This similarity indicates that the two wavelengths are mostly tracing the same dust component.   To determine the dust temperature and derive physical parameters we follow the technique presented in \citet{terebey09} and briefly described here. For thermal dust emission the intensity $I_{\nu}$ at frequency $\nu$ is given by  $I_{\nu}  =  \tau_{\nu} B_{\nu}(T)$, appropriate for low optical depth and where $B_{\nu}(T)$ is the Planck function and $\tau_{\nu}$ the optical depth. The wavelength dependence of the optical depth is given by the usual $\tau_{\nu} \sim \nu^{\beta}$ where $\beta =1.5 - 2$ at long wavelengths. If, for example, the dust temperature is constant across the image then the Planck term $B_{\nu}(T)$ is constant. The result is that the 160 and 100~\micron\ images will look identical, apart from resolution effects, while the structure in the image will linearly trace the optical depth i.e. column density. Also, a plot of I100 versus I160 intensity values for each image pixel will exhibit a linear trend whose slope is related to the dust temperature. In practice the assumption of constant dust temperature works well over scales of 1 to 2 degrees. 

Table~\ref{tbl:tdust} shows the I100 versus I160 slope and corresponding dust temperature for each of the seven OphN regions separately. There is a small but real trend in dust temperature across the region. The coldest dust temperatures are found in OphN~1 (15.6 K) and OphN~3S (15.9 K), located on the eastern side, while the warmest temperatures are in OphN~5 (17.0 K) and OphN~6 (16.8 K) located on the western side, nearest the OB association as seen in projection. The higher dust temperatures on the western side support the idea that the OB stars enhance the local interstellar radiation field. In comparison, the Taurus star-forming region has colder $14.2$~K dust (Terebey et al 2009, Flagey et al 2009), consistent with the lack of luminous heating sources in Taurus.

All but one of the Oph~N regions shows evidence for cold clumps in the excess map (Fig.~\ref{fig:excess}).  In OphN~1, these are associated with L260~SMM1 and 2.  The 160\micron\ excess peaks at the Class~1 protostar OphN~YSOc1 which is indicative of a cold disk or envelope.  By contrast,  there is no enhancement at the positions of the Class~II and III sources in OphN~3 although the clouds in the south of the region appear cold.  The effects of UV heating, presumably from $\zeta$~Oph, can be seen along the northern edges of these clouds as bright 100\micron\ features and negative values in the 160\micron\ excess maps.  OphN~4 exhibits one of the series of bright shells or loops best seen in the 100\micron\ maps.  The moderate extinction clump on the edge of this cavity is warm.  There is also little cold dust in OphN~5.  By contrast, OphN~6 shows two cold clumps and cold dust associated with the flat-spectrum source OphN~YSO8 (L1719B).

\section{Discussion}
\label{sect:discussion}

The Spitzer data advance our knowledge of Oph~N with an improved list of candidate YSOs, higher-resolution extinction maps, and dust temperature estimates.  In this discussion, we apply this information to revisit the YSO population, the current star formation rate/efficiency, and the evidence for triggering and other effects of the neighbouring Upper~Sco OB association on the region.

%Discussion of a few new interesting bits eg. clusters, low luminosity sources...
%MIPS: The most embedded objects
%Luminosity function of YSOs (eg. Harvey Serpens synthesis)
%Outflows (can be compared to H2)
% comparison with models?
\subsection{Star formation count and efficiency}

How many stars are currently forming in the Oph~N clouds?  In the areas covered by both \Spitzer\ IRAC+MIPS, ten YSO candidates are identified.  In the areas with no or limited IRAC coverage, MIPS+2MASS colours add a further five.  A further Class~I source, RNO91, is detected in L43 \citep{chen09}, bringing the total to 16.  This count almost doubles the nine sources known from IRAS \citep{carballo92}.   However, four of these candidates (including two of the IRAS detections) are probable AGB stars based on their colours and proper motions (Sect.~\ref{sect:agb}).  Excluding the AGB stars, the total number of \Spitzer-identified YSOcs in Oph~N is 12.

However, the area of Oph~N molecular clouds surveyed by \Spitzer\ is limited and it is possible that additional YSOs lie outside the mapped region.  Based on the low extinction these are likely to be Class~II/III sources.   We can estimate the number of YSOcs missed by \Spitzer\ by considering the number of IRAS detections within and outside the \Spitzer\ regions.   IRAS covers the whole of the Oph~N region but is less sensitive than \Spitzer: of the 12 \Spitzer\ YSOcs (including RNO91 in L43, excluding probable AGN), only seven have IRAS identifications.  

A wider area search for YSOs in Oph~N was made by \citet{carballo92} based on IRAS colours.  Excluding the Ophiuchus and Lupus clouds, they identify six IRAS sources in the Oph~N region as definite YSOs, five of which are cross-identified by \Spitzer\ and one of which lies outside the \Spitzer\ regions (T~Tauri stars V1003~Oph or RNO90 near L43).    A further \Spitzer\ YSOc (Oph~N YSOc3 or IRAS 16455-1405) is among a further ten IRAS sources ambiguously classified by \citet{carballo92} as either YSOs, galaxies or EGOs and one further \Spitzer\ candidate, YSOc~4, has an IRAS identification not listed by \citet{carballo92} (though included by \citealp{nozawa91}), bringing the total number of \Spitzer\ sources detected by IRAS to seven.  Hence, with the count of IRAS YSOs outside the \Spitzer\ regions numbering one, our \Spitzer\ survey encompasses 7/8 or 88\% of the IRAS-identified YSOs in the region.  

 In addition, there are four ambigous (as classified by \citealt{carballo92}) IRAS detections outside the \Spitzer\ regions which have associated extinction $A_V \geq 1$.  If these also turn out to be YSOs (eg. IRAS~16534$-$1557 or  IRAS~16439$-$0945; see sects.~\ref{sect:sco1} and \ref{sect:cb68}) the completeness decreases further to 7/12 or 58\%.  None of these are associated with MIPS-only sources.  Assuming these statistics hold for all 12 of the \Spitzer\ identified YSOcs, and not only the seven detected by IRAS, the true number of YSOs in Oph~N (north of the Ophiuchus L1688 and L1689 clouds) is likely to lie between 14 and 21.

The star formation surface density in Oph~N has been addressed recently by \citet{heiderman10} (under the region name `Scorpius'), who find a star formation rate per unit area $\Sigma_\mathrm{SFR} = 0.343 \pm 0.20\,\Msun \hbox{ yr}^{-1} \hbox{kpc}^{-2}$ based on the ten \Spitzer\ IRAC+MIPS identifications reported here and comparing with an $A_V>2$ area of $7.29\mathrm{ pc}^2$.  This is among the lowest in Gould belt clouds (mean $\Sigma_\mathrm{SFR} = 1.2\pm 0.2\,\Msun \hbox{ yr}^{-1} \hbox{kpc}^{-2}$), comparable to Lupus~I, Auriga~N and IC5146~E/NW but only 15\% of that in Ophiuchus ($2.45\pm 1.6\,\Msun \hbox{ yr}^{-1} \hbox{kpc}^{-2}$).  The inclusion of RNO91 and the three MIPS+WISE identifications, but the exclusion of two AGB stars, increases the rate to $\Sigma_\mathrm{SFR} = 0.412\pm 0.25\,\Msun \hbox{ yr}^{-1} \hbox{kpc}^{-2}$.   

An estimate of the instantaneous star formation efficiency (SFE) $M_* /(M_*+ M_\mathrm{gas})$ can be made comparing the total twelve YSOcs to the total mass in the Oph~N clouds, which is 3000\Msun\ calculated for $A_V>1$ in the \citet{dobashi05} map, %?? EXTINCTION CALCULATION FROM DOBASHI USES HEIDERMAN ET AL. CONVERSION?\\
consistent with 4400~\Msun\ from $^{13}$CO (\citealt{nozawa91}.  Assuming 12 YSOcs with an average stellar mass of 0.5~\Msun (following \citealt{evans09}) gives an overall SFE of 0.20\%.  Taking into account the survey incompleteness gives an upper limit of 0.34\%.  These efficiencies are consistent with the 0.3\% upper limit given by \citet{nozawa91} (we assume half their average stellar mass and a slightly different star count, 12 rather than 13). 

\citet{nozawa91} suggest two reasons for the low SFE: the cloud is clumpy and fragmented, with only a small fraction of the mass in high column density cores; and the UV radiation estimated from the nearby OB stars is a factor of 2 higher than the standard Habing field, leading to high cloud ionisation and magnetic support.  These estimates of the instantaneous SFE do not take into account the earlier formation of the main-sequence stars in Upper~Sco.  However, we confirm the result of \citet{nozawa91} that the current star formation efficiency in the cloud complex is very low. 

\subsection{Comparison with Ophiuchus}

This low star formation efficiency is completely unlike the nearby Ophiuchus~L1688 cloud, which is forming a dense embedded cluster \citep[e.g.][]{padgett08}.  Measured in low-excitation $^{13}$CO~1--0, the Oph~N and Ophiuchus cloud masses are similar:  4000\Msun\ compared to 3050\Msun\ (L1688, L1689 and L1712-1729 filament) \citep{loren89,nozawa91}.  The main difference between the cloud complexes is the mass in dense gas.  From the \citet{rowlesfroebrich09} 2MASS-based extinction maps, calculated identically for the two clouds, we calculate that Oph~N has 25\% the mass of Oph above $A_V=3$ (680\Msun\ in Oph~N), dropping to 4\% of the mass at $A_V>5$ and less than 0.5\% for $A_V$=10.
%Measured in $^{13}$CO~1--0, mean densities of clumps are lower in Scorpius, at $\log_{10}(\nhh /\hbox{cm}^{-3}) = 3.4$ compared to $3.7$ in the Ophiuchus clouds \citep{loren89}.  
\citet{tachihara00a} found that traced in C$^{18}$O~1--0, the cores in L1688 are ten times denser than the others in the region, and 32\% of the cloud mass is traced by C$^{18}$O compared to 8\% in the Oph~N clouds, indicating substantial optical depths in the $^{13}$CO\citep{tachihara00a}.  
As studies extend beyond the more obvious star-forming regions, many molecular clouds are seen to fit this pattern of extended low column density material with a few low-mass cores, but the Oph~N clouds are particularly inefficient:  the Lupus molecular clouds -- also on the edge of Sco-Cen association -- have a star formation efficiency of a few percent \citep{tachihara96,merin08}; Taurus is also less extreme than Oph~N with a star formation efficiency of 1--2\% \citep{mizuno95}.  Maddalena's cloud (G216-2.5) is a more extreme example with a mass of more than a million \Msun\ only a few dozen YSOs \citep{megeath09}.

\subsection{Prestellar cores}

There is no evidence that star formation is coming to an end in the Oph~N region.  The YSO candidates cover the full range of evolutionary classes (3 Class 0/1; 1 flat-spectrum; 7 Class~II; 1 Class~III), and there are also several dense starless cores in the region, including submm cores in L234E, L255~SMM2 in OphN~1, and L158 in OphN~3a.  Star formation is clearly still ongoing in Oph~N, though at a low rate.
 
A statistical estimate of the starless core lifetime can be made from the high extinction cores.  Considering $A_V>10$ cores, the ratio of starless to protostellar cores is 12:3 (from Fig.\ref{fig:extinction}, including CB68, L234E and one starless and one protostellar core in L43).  If the starless cores are prestellar cores, then this ratio suggests their lifetime is a factor of 4 longer than the protostellar lifetime of 0.54~Myr \citep{evans09,class} or $\sim 2$~Myr.  This is consistent with expectations for lifetime vs. volume density \citep[][ Fig.2] {wt07} given a typical core density of $7\times 10^3\hbox{ cm}^{-3}$ \citep[][ from C$^{18}$O]{nozawa91}.  This is further support for the idea that the restrictive step in Oph~N is the formation of dense cores from low density material.

\subsection{Influence of Upper~Sco OB association}

With a known age gradient through the OB associations from Upper Centaurus-Lupus to Upper~Scorpius to Ophiuchus, Ophiuchus~North would seem an ideal region in which to look for the effect of triggering on star formation.  But while there is ample evidence for the dynamical and radiative influence of Upper~Sco on the region, the low star formation efficiency and normal fraction of starless cores do not suggest that star formation has been promoted, even in the clouds most strongly affected by the nearby OB association.

An expanding HI shell surrounds Upper~Sco,  created by the stellar winds and possibly the supernova explosion of the binary partner of runaway $\zeta$~Oph, impacting on the Ophiuchus and Oph~N clouds from the west \citep{degeus92}.  According to \citet{degeus92}, the expansion of the shell in this region is halted by the Ophiuchus and OphN~6 dense cores, with molecular gas swept from the main clump to produce OphN~4-5 and the elongated Ophiuchus filaments to the south.  A relatively recent origin is consistent with the absence of star formation in OphN~4 and 5 but could be tested by chemical comparison with OphN~6.  Both OphN~5 and OphN~6 have elevated temperatures (Sect.~\ref{sect:long}).  The bubbles seen in the 100\micron\ maps of OphN~4 and 5 also suggest dynamic disruption of these clouds.   A similar though less extreme lack of star formation is seen in Oph~L1689 compared to L1688,  where the effects of the O-star $\sigma$~Sco are strongest \citep{nutter06}.  The Oph~N~6 and Ophiuchus clouds are at similar distances to Upper~Sco, so the physical input from winds and radiation must have been similar in each region.  The difference in the star formation points to differences in the initial conditions, with the gas in Oph~L1688 sufficiently gravitationally bound to survive - and even benefit from - the onslaught from Upper~Sco, whereas the gas in OphN~6 had too low a density to counter the feedback forces.  Timescales are too short for the formation of the Class~II and III sources in the L43, OphN~3 and 6 clouds to have been initiated by this most recent shock, which has an upper limit on its expansion timescale of 2.5~Myr and may not even yet have reached these clouds \citep{tachihara96}.  There is also no clear sequence in protostellar ages from the front to the back of the filament which would suggest that the star formation has been triggered by the interactions with Upper~Sco OB association.  

The UV field from the runaway O-star $\zeta$~Oph forming the S27 H{\sc II} region is possibly an earlier influence on the northern Oph~N clouds and L43 \citep{tachihara00b}.  The dense gas in OphN~1--2 mainly lies along the west side of the filament and the clouds are also heated along this side (Sect.~\ref{sect:long}).  The 100\micron\ maps show bright rims to the west and north in the direction of $\zeta$~Oph and sharp boundaries in the dust distribution.  $\zeta$~Oph (09.5) originates in Upper Sco and has since tracked from southwest to northeast across the northern Oph~N region on timescales of a few $\times 10^5$ years \citep[see][Fig.~5]{tachihara00b}.  The radiation has impacted OphN~1 and L43  and may have contributed to the compression of the cores hosting Class~0/I sources in L43 and L260.   The 2~Myr old T~Tauri stars in OphN~3 and RNO90 near L43 predate the impact of the H{\sc II} region.  A study of the energetics suggests that UV photodissociation from $\zeta$~Oph is currently accelerating gas from the remaining dense cores towards the east \citep{tachihara00b}.  This is also suggested by the bright rims and cometary nature of OphN~1--3 as seen in the 100\micron\ IRAS map (Fig.~\ref{fig:long}).  It seems likely that the small star-forming molecular cores in Oph~N are the surviving remnants of the giant molecular cloud which produced Upper~Sco and are now slowly being ablated by the massive stars.

%Tachihara et al: only difference is that star-forming cores generally have higher NH2, and highly elongated cores haven't formed stars.  Column density threshold is higher than in Taurus, 8 x 10^21 cm^-2.  Suggest another effect, eg. stronger coupling with the magnetic fields, might prevent cores from collapsing.

%Goodman et al. 1994 Zeeman 10 microgauss, 1990 optical polarimetry

%Why Class 0s in Sco~1 and CB68 (and Oph) and 17 in L1688?
%Compare mass of cores w/wo Class Is
%Magnetic field orientation?  Check velocity
%Local gravitational field - Goodman nature paper? 

% Extended emission (PAH emission in 8 micron band, MIPS 160 micron)
% Shock excited gas (4.5 micron H2)

\section{Conclusions}
\label{sect:summary}

We have surveyed the high extinction regions ($A_V > 3$) of the Oph~N molecular cloud complex in the mid-infrared with \Spitzer\ IRAC and MIPS.  

There is little active star formation in Oph~N.  Twelve YSOcs are identified in total.  Eight of these are identified from their IRAC-MIPS colours.   An additional Class~I protostar exists in the L43 cloud in the centre of this region \citep{chen09}.  Three further YSO candidates are found based on red $[K-24]$ colours.  Four sources initially selected as YSO candidates were rejected as AGB stars and one as a galaxy.  Of the twelve remaining YSOcs, three are Class~I, one flat-spectrum, seven Class~II and one Class~III.   All except one Class~III and two Class~II sources were previously known.   The region as a whole has a low star formation efficiency of $<0.34\%$.

The high extinction regions in Oph~N are fragmented into twelve small ($\sim 0.2$~pc), scattered and low-mass ($\sim 10\Msun$ or less) cores, of the kind which might form one or two low-mass stars.  Three of these (OphN~1, CB68 and L43) currently contain Class~I sources.  The remainder are starless.

The interstellar medium in this region is influenced by the Upper~Sco subgroup of the Sco-Cen OB association.  There is evidence for dynamical interaction in the OphN~4 and OphN~5 bubbles, elevated temperatures in OphN~5 and 6, and irradiated cloud edges throughout the region.  The bulk of the gas mass is at low column density with only a few low-mass cores surviving to form a few YSOs.  This is very different from the situation in nearby Ophiuchus~L1688, which contains hundreds of YSOs \citep{padgett08}, but similar to the Lupus clouds \citep{tachihara96,merin08}.  As the impact of radiation and winds from Upper~Sco must have been similar in both regions, the initial conditions must have been very different.   Whereas the main Ophiuchus complex contained hundreds of solar masses of dense gas, Oph~N only contained a few low-mass high-density cores.  It seems likely that the small star-forming molecular cores in Oph~N are the surviving remnants of the GMC which produced Upper~Sco and are now slowly being ablated by the massive stars.  The low star formation rate is due to the lack of dense cores.

\section*{Acknowledgments}

Our thanks to the anonymous referee for their constructive suggestions.  ST would like to thank Deborah Padgett for help with the WISE data.  JH acknowledges support from the STFC Advanced
Fellowship programme.   EEM acknowledges support from NSF
award AST-1008908.  This research has made use of the SIMBAD database,
operated at CDS, Strasbourg, France; the NASA/ IPAC Infrared Science Archive, which is operated by the Jet Propulsion Laboratory, California Institute of Technology, under contract with the National Aeronautics and Space Administration; and data products from the Wide-field Infrared Survey Explorer, which is a joint project of the University of California, Los Angeles, and the Jet Propulsion Laboratory/California Institute of Technology, funded by the National Aeronautics and Space Administration. 
\bibliographystyle{apj}
\bibliography{ophN_v2}

\appendix
\section{Comments on individual regions}
\label{sect:regions}

\subsection{OphN~1}
\label{sect:sco1}

The two extinction peaks in OphN~1 lie in Lynds opacity class 6 cloud L260  (Fig.~\ref{fig:extinction}).  The YSOc and starless core are both in the more massive northern extinction region (9 \Msun\ above an extinction limit of $A_V > 10$, compared to a C$^{18}$O masses of 11.4\Msun;\citealp{nozawa91}).   The southern extinction peak shows no evidence for star formation or dense cores and little high-extinction gas (1\Msun\ above $A_V = 10$), indicating that the C$^{18}$O mass of 17.9\Msun\ mainly lies at lower extinctions.

The Class~I source YSOc1 is identified with IRAS~16442-0930/IRAS~16449-1001 and lies close to the peak of the northern core (\citealt{tachihara00a} C$^{18}$O core u2).  This YSOc was first identified as such from the IRAS catalogue by \citet{carballo92}.  The source is detected but unresolved in 1.3mm continuum with a mass $<0.05$~\Msun\ and at 850\micron\ by SCUBA with a peak flux less than $72$~mJy/beam \citep[L260~ SMM1][see our Fig.~\ref{fig:scuba}]{visser02}, consistent with the mm/submm emission largely originating from a disk.  There is an associated $^{12}$CO outflow \citep{batc96,visser02} and a H$_2$O maser \citep{han98}.   The IRAS spectrum of this source was studied in detail by \citet{parker91}, who concluded from a shoulder in the short-wavelength emission at 1.5--2\micron\ that it was a Class~II YSO obscured by an extended dark cloud, but we see no evidence for this shoulder in the 2MASS fluxes.  

The submm emission and the extinction in this region peaks towards a second core $3'$ to the east \citep[][ L255 SMM2]{visser02}.  No source is detected by \Spitzer\ even at 70\micron\ (Fig.~\ref{fig:scuba}), confirming this as a starless core.

A second IRAS source IRAS~16454-0955 with ambiguous colours typical of a YSO or EGO lies in the southeast of the IRAC region unobserved by MIPS \citep{carballo92} (marked in Fig.~\ref{fig:scuba}).  A search through the positional matches within $10''$ in the \Spitzer\ catalogue found no reddened or rising spectra.  Better waveband coverage is needed to pinpoint the source of the infrared emission detected by IRAS.  Another ambiguous IRAS YSO or galaxy from \citet{carballo92}, IRAS~16439-0945, lies to the west on the edge of the OphN~1 $A_V \sim 2$ region, but off the IRAC and MIPS maps.

%OTHER MIPS1 SOURCES: IRAS 16445-0942, BD-09 4450, HD 151503, 2MASS J16471185-0928363, BD-09 4452, BD-10 4391, HIP 82197.  These are not YSO candidates.

\subsection{L234E}

South of OphN~1, L234 (Lynds opacity class 3) contains no YSOcs.  The area mapped by \Spitzer\ has a mass of only 1\Msun\ above $A_V>10$, though it is coincident with \citet{tachihara00a} core t (C$^{18}$O  mass 12.9\Msun).  The SCUBA map shows weak 850\micron\ emission peaking at 130~mJy/beam \citep{kirk05}.  The faint 70\micron\ source seen an arcminute to the north in Fig.~\ref{fig:scuba}, at $16^\mathrm{h}48^\mathrm{m}06\fs6 -10\degr56\arcmin06\arcsec$, is classified by the c2d pipeline as a galaxy.

\subsection{OphN~2}

This region in L204 continues the filament of bright-rimmed clouds, and contains two extinction peaks over $A_V=10$ (8 and 10\Msun respectively), several C$^{18}$O cores (\citealt{tachihara00a} s1--5) with masses from 5--33\Msun, but no YSOcs.  No submm maps exist of this region.  It is a good candidate for future star formation.

\subsection{OphN~3}
\label{sect:sco3}

%2  &16:48:17.6&-14:11:09.1  &Sco3 &II &V* V2507 Oph (8.7'') & &D05\\
%3  &16:48:21.9&-14:10:42.8  &Sco3 &II &IRAS 16455-1405 (0.27'') & &VRC02,BATC96 (Class I),CWW92\\ %L158, see Visser 2002
%4  &16:48:45.6&-14:16:36.1  &Sco3 &II &V* V2508 Oph (0.16'') &L162 SMM1$^{1}$\\
%5  &16:49:00.8&-14:17:10.9  &Sco3 &II &Oph~5 (0.12'')\\
%6  &16:49:05.6&-15:37:13.1  &Sco3 &III &IRAS 16462-1532 (7.45'')\\

Five Class~II and III protostars are identified in the OphN~3 IRAC fields, four in the North and one in the South field.  The northern sources are associated with Lynds opacity class 6 cloud L162 but are not coincident with the extinction peaks: they lie on the edge of the $A_{\mathrm v}=3$ region heated by $\zeta$~Oph
%away from the C$^{18}$O peak r2 \citep{tachihara00a}, 
(see Fig.~\ref{fig:extinction}).  One of the northern sources, OphN~YSOc3, is associated with IRAS~16455$-$1405, already identified as a YSO or (ambiguously) an EGO by \citet{carballo92}.  This star is in fact a binary with $1''$ separation (Reipurth~44; \citealp{reipurth93}) and possibly even a triple system \citep{koresko02}.    Another potential binary is OphN~YSOc2, which lies $9''$ from the extended red variable T~Tauri star V*~V2507~Oph; both have similar proper motions consistent with the idea that these two stars form binary Reipurth~43 \citep{reipurth93}.

The classical T~Tauri star V* V1121 Oph (IRAS 16464-1464)  lies 10 arcminutes to the southeast of the other northern OphN~3 YSOcs at $16^{\mathrm h} 49^\mathrm{m} 15\fs3 -14\degr22\arcmin08\farcs6$ \citep[][see our Fig.~\ref{fig:scuba}]{am94};.  It is detected by IRAC bands 1 and 3 but lies outside the field of IRAC 2 and 4 and the YSOc identification comes from the $K_S-[24]$ selection (Table~\ref{tbl:mipsysocs}).  At 24 and 70 microns it is bright but extended and confused with cirrus and the catalogue fluxes are bandfilled (see Figs.~\ref{fig:scuba} and \ref{fig:extinction}).  V1121~Oph is classified as a Class~II source \citet{am94} and, like the other Class~II YSOcs 4 and 5 in OphN~3, it is bright at 70\micron\ (Fig.~\ref{fig:scuba}). 

The \citet{visser02} SCUBA 850\micron\ map of the L158 and L162 regions is overlaid on the 24\micron\ map in Fig.~\ref{fig:scuba}.  YSOc4 (Class~II) is detected as a compact 850\micron\ source L162~SMM1 which \citet{visser02} identify with  IRAS~16459$-$1411.  This is a well-studied T~Tauri star with a disk \citep{am94,andrews09}.  YSOc5 is not detected by SCUBA at a level of 100~mJy/beam.  Neither YSOc2 or YSOc3 lie within the SCUBA maps.  The MIPS-only Class~III source M1 which lies in the southeast of the 70\micron\ map in Fig.~\ref{fig:extinction}  is identified as IRAS~16484$-$1557.  The starless core in L158 detected by SCUBA \citep{visser02} remains infrared-dark even at 70\micron, confirming this source as starless.  The nearby IRAS sources IRAS16445$-$1352, IRAS~16442$-$1351 and IRAS~16439-1353 appear to be cirrus.  The L158 core lies in the most massive of the three northern extinction peaks, which have $A_V > 3$ masses of 5, 1 and 13\Msun, measuring from east to west.  

The galaxy 6dFGS~gJ164828.8-141437 confusingly lies among the northern OphN~3 YSOcs, and was initially identified by the colour-colour YSOc selection procedure but rejected when examined by eye as being elliptical.  It appears in the 2MASS-MIPS colour-magnitude diagram (Fig.~\ref{fig:kmips}) marked as a YSOc but clustered with the galaxies at $K_S \sim 14$.  Its location within $20'$ of the other YSOcs in the north of OphN~3 is particularly confusing (see Fig.~\ref{fig:scuba}).

%Several IRAS sources appear to be cirrus: IRAS~16455-1415, IRAS~16457-1409, IRAS~16464-1407.
%IRAS~16453-1408 shows a strong 24micron source but has a falling spectrum. 

In the southern region we identify OphN~YSOc5, classified Class~III, with IRAS~16462$-$1532.  This source is associated with the westernmost of the two extinction peaks, mass 13\Msun\ ($A_V > 10$; see Fig.~\ref{fig:extinction}).  The southeastern extinction core is more massive, at 21\Msun, and a likely site for future star formation.  There are no submillimetre maps of this region to determine if it contains any dense cores.  The Lynds opacity 4 and 5 dark clouds L137 and L141 lie to the north of this region, but are not associated with any YSOcs.

%V* V1010 Oph NOT DETECTED AS A YSO - eclipsing binary  16 49 27.67 -15 40 04.69

\subsection{OphN~4}

The OphN~4 field is associated with L1757, which shows no extinction above $A_V=10$, consistent with its detection in $^{13}$CO \citep[][ cloud C]{nozawa91} but not C$^{18}$O \citep{tachihara00a}.  There is a bright-rimmed globule in the 24\micron\ map but no YSOcs.   Ambiguous YSO/EGO IRAS~16355-1807 is located at the northern end of the extinction filament 2~degrees east of OphN~4, at an optical extinction of $A_V\simeq 1$ \citep{carballo92}. 

\subsection{OphN~5}

The OphN~5 field contains no mass above $A_V > 10$ but no YSOcs.  The dark cloud L1752 is centred further to the east.  OphN~5 is also associated with $^{13}$CO cloud C \citep{nozawa91} but no C$^{18}$O core \citep{tachihara00a}.

% IRAS16249-1919 has a stellar 24 micron detection, but only limits on IRAS 25-100 and is associated with extinction region - star? 16 24 54.6 -19 19 31| 46 10  99|    0.38     0.46L    1.98L   22.62L

\subsection{OphN~6}
\label{sect:sco6}

% Matches use Gaia to overlay IRAS catalogue
% 7  &16:21:37.7&-20:00:36.9  &Sco6 &III &IRAS 16186-1953&\\
% 8  &16:22:04.3&-19:43:26.8  &Sco6 &F &IRAS 16191-1936 (1.04'')&\\
% 9  &16:22:09.6&-19:53:00.9  &Sco6 &III &--&\\

OphN~6 is associated with the L1719 dark cloud and two high extinction regions, \citet{tachihara00a} C$^{18}$O cores a (south) and b (north), and contains two Class~III YSO candidates and one flat-spectrum source (YSOc8).  YSOc7, associated with IRAS~16186-1953,  is probably a background giant due to its proper motions (see Sect.~\ref{sect:agb}).  YSOc7 is variable in the optical/near-infrared; the \citet{denis05} reports variation of more than a magnitude in $I$ and $J$ bands.  YSOc8 is associated with IRAS~16191-1936.  OphN~YSOc9, has a small 24\micron\ excess and was previously unknown.  The YSOcs lie away from the extinction peaks, but two lie on the edges of high extinction regions: Class~III YSOc7/IRAS~16186-1953 in the south, and the flat-spectrum YSOc8 at the edge of the northern extinction peak.  

The flat-spectrum source YSOc8 is associated with L1719B and has a CO outflow \citep{batc96}.  The submillimetre emission is still uncertain: while there is a 220 mJy detection at 1.3mm suggesting a disk mass of 0.04 Msun \citep{am94,andrewswilliams07}, SCUBA observations only give an 850\micron\ upper limit of 54 mJy/beam \citep{kirk05}.  Weak submillimetre emission was detected with SCUBA towards L1719D at $16^{\mathrm h}21^\mathrm{m}09\fs2 -20\degr 07\arcmin10\arcsec$ with a flux of 62~mJy/beam.  These weak emission SCUBA jiggle maps do not reproduce well as contours and are not shown in Fig.~\ref{fig:scuba}; see \citet{kirk05} for images.

OphN~6 hosts two of the MIPS+2MASS identified YSOcs: M3 and M4.  Both of these have proper motions consistent with Ophiuchus distances \citep{PPMXL}.  M4 lies $9''$ away and within the positional errors of ROSAT X-ray source 2RXP~J162230.1-200256, supporting its identification as a YSO.

The bright region in the southwest of OphN~6 contains four late-type B/early A stars: HIP~80019, HIP~80024 and HIP~80062/3.  The revised Hipparcos parallaxes \citep{vanleeuwen07} and Tycho-2 proper motions for these stars are given in Table~\ref{tbl:bstars}.  The weighted mean of the four parallaxes gives a distance of $164^{+18}_{-15}$~pc, larger than our assumed 130~pc (equivalent to a parallax of $7.4$~milliarcseconds) but consistent with parallax distances for other stars illuminating nebulae in the vicinity of Sco/Oph \citep{mamajek08} and with distance estimates for Upper~Sco ($145\pm2$~pc; \citealt{dezeeuw99}).  The proper motions are also similar to each other and consistent with membership of Upper~Sco \citep{mamajek08,hoogerwerf00}.  Three of these stars are clearly associated with the cloud through nebulae (see the 3-colour images and 24~\micron\ emission, Figs.~\ref{fig:rgbirac}, \ref{fig:rgbmips} and \ref{fig:scuba}).  HIP~80019 has no apparent nebula.   HIP~80062/3, a wide B9 binary with HIP~80063 itself a close binary, illuminates the optically-visible reflection nebula IC4601 \citep{magakian03}.   

It is not yet clear if these stars are pre-main-sequence.  None of them have the red colours characteristic of YSOs with disks (Fig.~\ref{fig:ysosel} and \ref{fig:kmips}).  Neither HIP~80019 nor 80024 show emission lines characteristic of Herbig~AeBe stars \citep{hernandez05}.  However, \citet{dahmcarpenter09} found a small excess above 10\micron\  for HIP~80024 in \Spitzer\ spectroscopy and characterised this source as a debris disk with the proviso that they `cannot rule out that the dust is remnant primordial material'.  This star lies outside the region mapped at 24\micron\ in our study so we can neither confirm or deny this disk excess.  Now that their revised Hipparcos parallaxes and Tycho-2 proper motions are consistent with membership of Upper~Sco, the two stars HIP~80062/63 driving IC4601 would be interesting targets for further studies.

The bright star saturated in the IRAC bands at $16^\mathrm{h}20^\mathrm{m}20\fs9  -20\degr08\arcmin05\arcsec$ is identified as 2MASS~J16202094$-$2008059, and also saturated in 2MASS $H$ and $K_S$.  It is not obviously associated with any local nebulosity and, given its extremely red colours ($V-K \sim 10$, \citealp{nomad}), is probably a background giant. 

%which is A/B-type exciting sources (numbers 102 and 103a/b in the catalogue of \citealt{vandenbergh66}, Hipparcos %stars HIP~80062/80063 in \citealt{vanleeuwen07}).  The reflection nebula lies on the fringes of the Upper~Sco association %to the northwest.  The B-stars .  
% t, and proper motions of 40~mas/year to the East and South also indicate a common origin with Upper~Sco (ref?
% Other IRAS sources are cirrus except 16174-2001 which is a bright 24micron star with rising IRAS fluxes  1.97     0.59:    8.73:   24.30

\subsection{CB68}
\label{sect:cb68}

The CB68 globule (L146) contains a well-studied isolated Class~0 object, from the IRAS faint source catalogue IRAS 16544$-$1604.  The associated submm emission is bright and extended $40''$ in size with peak flux densities of 2.1 and 0.42~Jy/beam at 450 and 850\micron\ respectively and an estimated mass of 0.1\Msun\ \citep{huard99,vallee03,young06}.  The molecular core is elongated northeast-southwest, with a perpendicular molecular outflow detected in $^{12}$CO \citep{vallee00,vallee03}.  The dust emission shows strong (11\%) polarisation yielding an estimated magnetic field strength of 120--130~$\mu$G in the radial direction \citet{vallee07}.  CB68 also has a C$^{18}$O detection (\citealt{nozawa91} $^{13}$CO core~34) though the \citet{tachihara00a} C$^{18}$O core q2 lies 0.4~degrees further to the north.

\begin{figure}
\epsscale{1.0}
%\plottwo{PS/Sco_SEDs_all_0_bw.eps}{PS/Sco_SEDs_all_0.eps} %original from IDL.  GIMP BW.
%\plotone{PS/Sco_SEDs_all_0.eps} %original from IDL.  GIMP BW.
\plotone{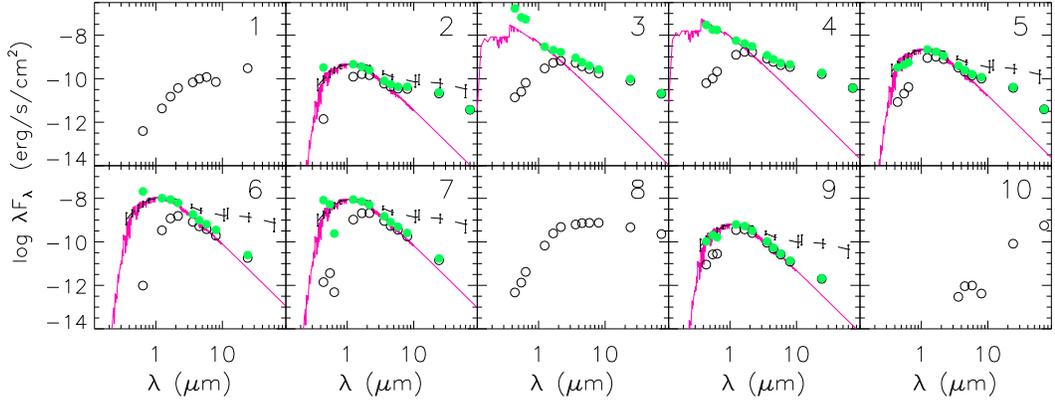} %original from IDL.  GIMP BW.
\caption{Spectral energy distributions for the ten YSOcs listed in Table~\protect\ref{tbl:ysos}.  Measured \jh{NOMAD}, 2MASS, IRAC and MIPS fluxes are shown as open circles.  For the evolved sources (flat-spectrum, Class II and Class III), photospheric models are plotted for a K7 star except for Sources 3 and 4, for which an A0 photosphere was used.  The filled circles (green in the online version) then show the fluxes corrected for reddening.  For the Class~I sources 1 and 8, no photospheric model is plotted.  The dashed line shows the median SED of T~Tauri stars in Taurus (with errorbars showing the quartiles of the distribution) \protect\citep{hartmann05}.  Uncertainties are smaller than the symbol size (see Table~\ref{tbl:fluxes}). \label{fig:seds}}
\end{figure}

\begin{figure}
\epsscale{0.5}
%\plotone{PS/kmips-all-v2-crop.eps} %original from python
\plotone{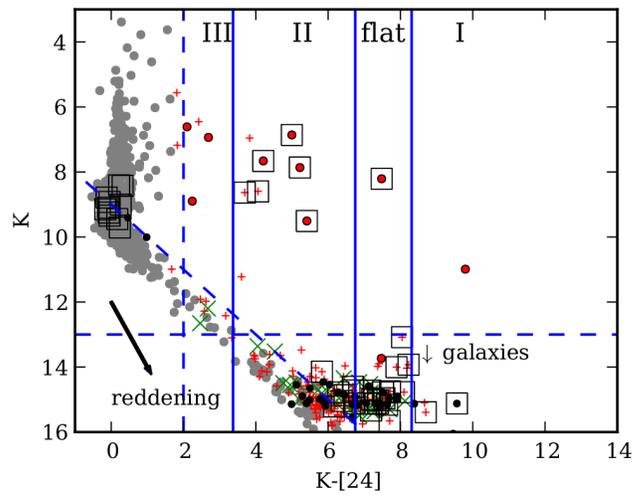} %original from python
 \caption{$K_S$ vs. $K_S-[24]$ colour-magnitude diagrams for all Oph~N regions combined.  Symbols reflect the c2d pipeline classification: stars (grey dots); star+dust (red plus symbols in the online version); galaxies ($\times$, green in the online version); YSOc (black circles, filled red in the online version); and other (black dots).  Class~III, Class~II, flat-spectrum and Class~I source models and the horizontal and diagonal lines mark cuts for faint $\mathrm K_S$ and 24\micron\ detections. \label{fig:kmips}}
\end{figure}

\begin{figure}
\epsscale{1.0}
%\plottwo{PS/kmips_seds_bw.eps}{PS/kmips_seds.eps} % IDL
%\plotone{PS/plotwiseseds.eps}
\plotone{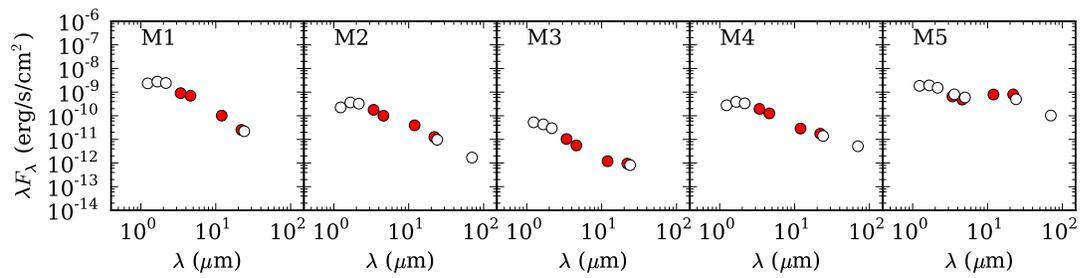}
\caption{Spectral energy distributions for sources detected via K-24\micron\ colours (see Table~\protect\ref{tbl:mipsysocs}).  Fluxes from 2MASS and \Spitzer\ are plotted as open circles and those from WISE as filled circles (red in the online version).}
\label{fig:kmipsseds}
\end{figure}

\begin{figure}
\epsscale{0.3}
%\plotone{PS/agbcol.eps} % original eps from python.  B/W required.
\plotone{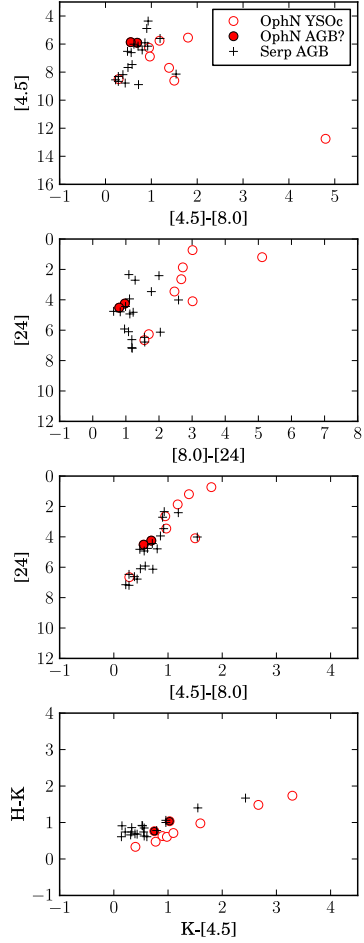} % original eps from python.  B/W required.
\caption{Colour-magnitude diagrams for the IRAC-detected YSOc (open circles) comparing the AGB candidates (filled circles) with the spectroscopically identified AGB stars in Serpens (crosses, Harvey 2012 priv. comm., from \citealp{oliveira09}).  The color-magnitude combinations are the same as Fig.~\ref{fig:ysosel}, though unlike that figure the galaxy interloper 6dFGS gJ164828.8 -141437 is not shown.  The YSOc occupying the same region of the graph as the AGB stars, but with the lowest 24\micron\ fluxes and bluest $[4.5]-[8.0]$ colours, is YSOc~9. \label{fig:agbcol}}
\end{figure}

\begin{figure}
\epsscale{1.0}
%\plottwo*{PS/mips2+scuba_bw.eps}{PS/mips2+scuba.eps} % converted with EPS.  
%\plotone{PS/mips2+scuba.eps}
\plotone{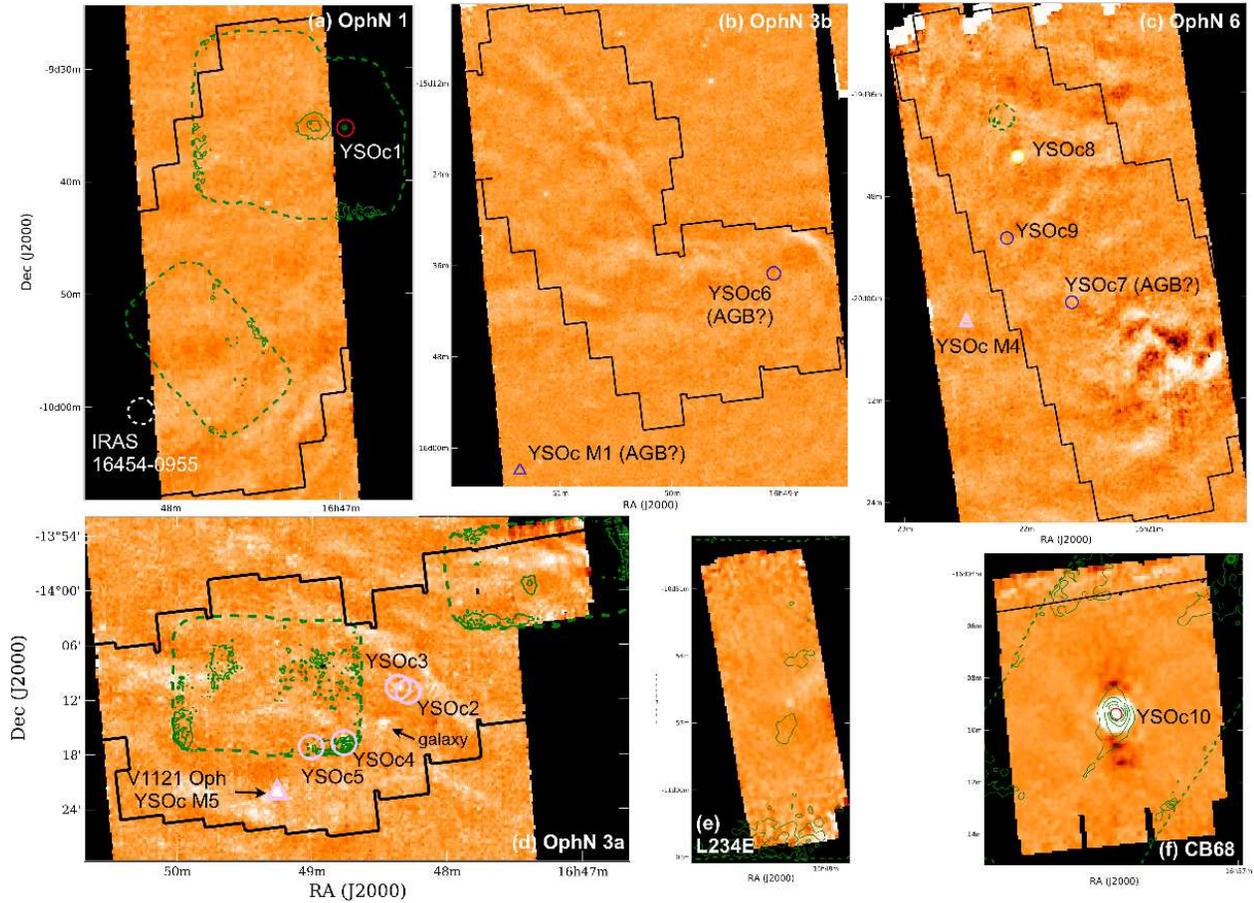}
\caption{SCUBA 850\micron\ contours (green in online version, from 100~mJy/beam in steps of 100~mJy/beam) overlaid on MIPS 70\micron\ maps (grey/colourscale, logarithmic from $-10$ to $5$~MJy/sterad).  Top to bottom, left to right: ({\em a}) OphN~1, ({\em b}) OphN~3b, ({\em c}) OphN~6, ({\em d}) OphN~3a, ({\em e}) L234E and ({\em f}) CB68.  The regions mapped by IRAC (band 1) are marked in black.   Circles and triangles represent IRAC-detected and MIPS-only YSO candidates from Tables~\protect\ref{tbl:ysos} and \protect\ref{tbl:mipsysocs} respectively.  YSOcs identified using IRAC and MIPS are marked with circles,and those identified using MIPS and 2MASS with triangles (in the online version, coloured according to YSO class: red=Class~I, yellow=flat-spectrum; pink=Class~II; blue=Class~III).  The T~Tauri star V1121~Oph,  the confusing galaxy gJ164828.8 -141437, and the possible YSO IRAS~16454-0955 are also marked.  The three regions which are starless and unobserved by SCUBA are not shown (OphN~2,4,5).}
\label{fig:scuba}
\end{figure}

\begin{figure}
%\epsscale{0.5}
%\plottwo{PS/mips24+av.eps}{PS/mips24+av_bw.eps} % GIMP using Color Tools -> Desaturate for BW
%\plotone{PS/mips24+av.eps}
\plotone{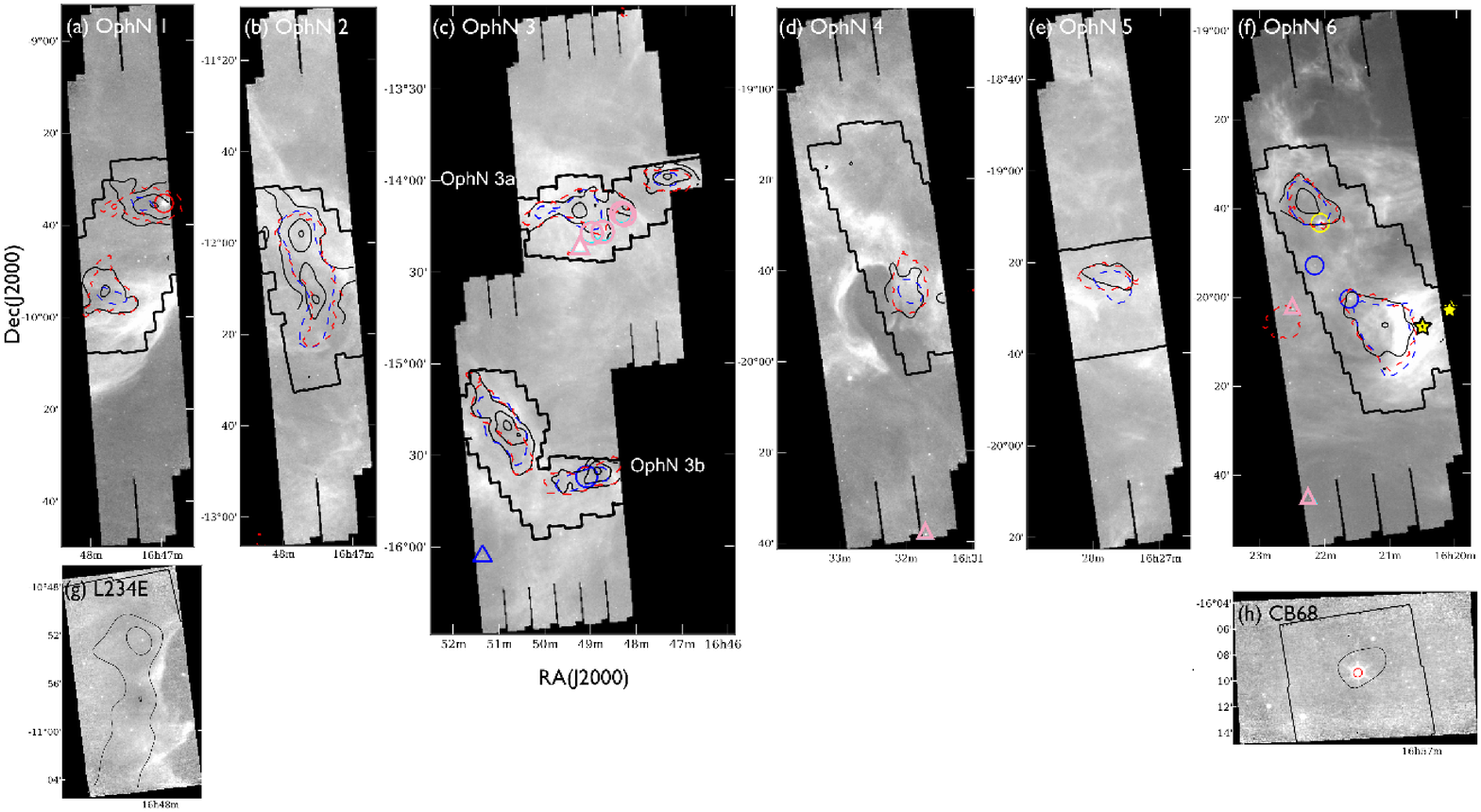} % new version for resubmission, colour and BW

\caption{MIPS 24\micron\ maps (colourscale) overlaid with Spitzer-derived extinction contours at $A_V = [6,10,20]$ (black) for the areas observed by IRAC.   Top left to bottom right: ({\em a--f}) OphN~1,2,3,4,5 and 6, ({\em g}) L234E and ({\em h}) CB68.  For comparison, the $A_V=3$ level from optical starcounts \protect\citep{dobashi05}  and $A_V=4$ from near-infrared colours \protect\citep{rowlesfroebrich09} are plotted with dashed contours (blue and red respectively in the online version). Sources are marked as in Fig.~\protect\ref{fig:scuba}.}
\label{fig:extinction}
\end{figure}

\begin{figure}
%\epsscale{0.7}
%\plottwo{PS/100and160.eps}{PS/100and160_bw.eps} % GIMP
%\plotone{PS/100and160.eps}
\plotone{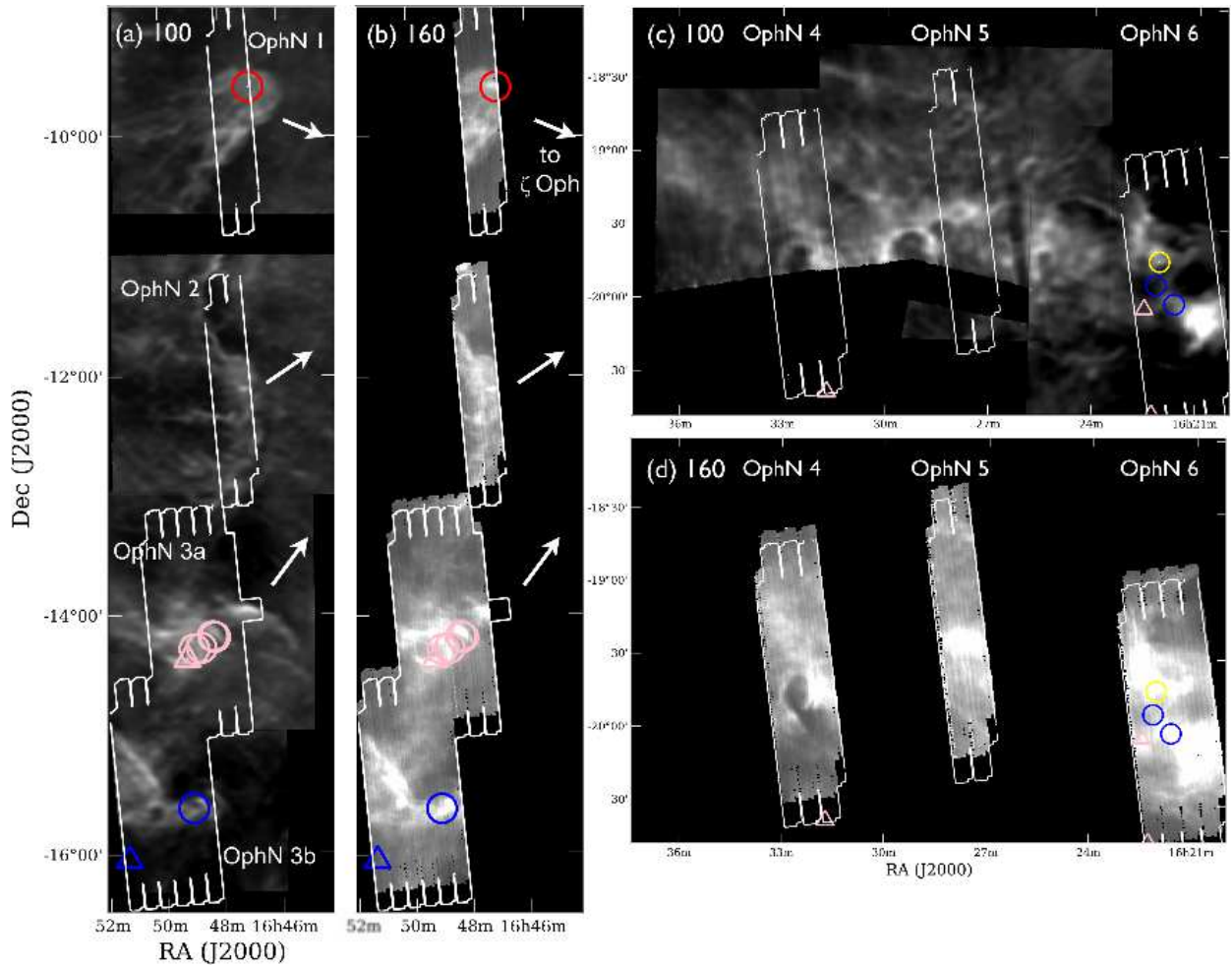} % colour and BW
\caption{Spitzer 160\micron\ emission and IRAS HIRES-processed 100\micron\ emission. Left to right, top to bottom: ({\em a}) IRAS 100\micron\ image of OphN~1,2, and 3, with arrows indicating the approximate direction of $\zeta$~Oph; ({\em b}) \Spitzer\ 160\micron\ image of OphN~1,2 and 3; ({\em c}) IRAS 100\micron\ image of OphN~4,5 and 6; ({\em d}) \Spitzer\ 160\micron\ image of OphN~4,5 and 6.  Greyscales are a square-root stretch from -10 to 50 MJy/steradian (IRAS 100\micron) and 10 to 150 MJy/steradian (\Spitzer\ 160\micron).  Sources are marked as in Fig.~\protect\ref{fig:scuba}.}
\label{fig:long}
\end{figure}

\begin{figure}
%\epsscale{0.5}
%\plottwo{PS/160excess.eps}{PS/160excess_bw.eps} % GIMP
%\plotone{PS/160excess.eps}
\plotone{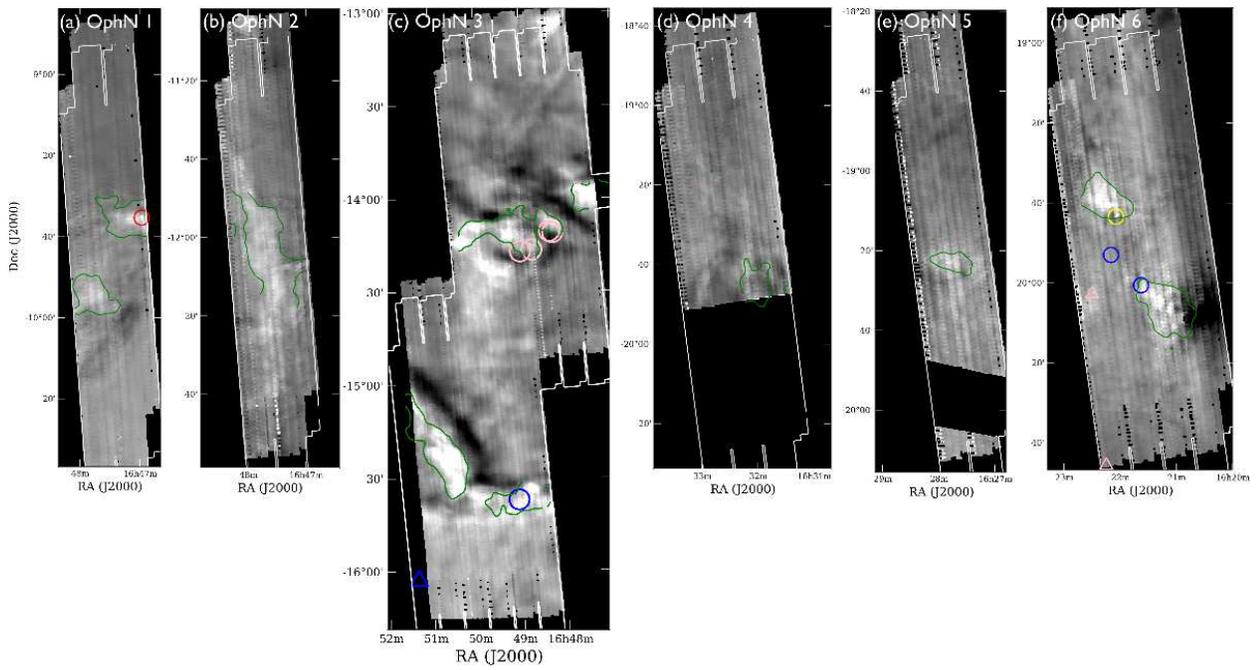}
\caption{Excess 160\micron\ over 100\micron\ emission (greyscale, linear from -50 to 50 mJy/steradian) overlaid with Spitzer $A_V=6$ contour (green in the online version, shown in the IRAC regions only).  Left to right, ({\em a--f}), regions OphN~1, 2, 3, 4, 5 and 6.  CB68 and L234E have no 160\micron\ map and are not shown.  Sources are marked as in Fig.~\protect\ref{fig:scuba}.}
\label{fig:excess}
\end{figure}

\begin{figure}
\epsscale{0.65}
%\plottwo{PS/cb68_irac2_zoom.eps}{PS/cb68_irac2_zoom_bw.eps}
%\plotone{PS/cb68_irac2_zoom.eps}
\plotone{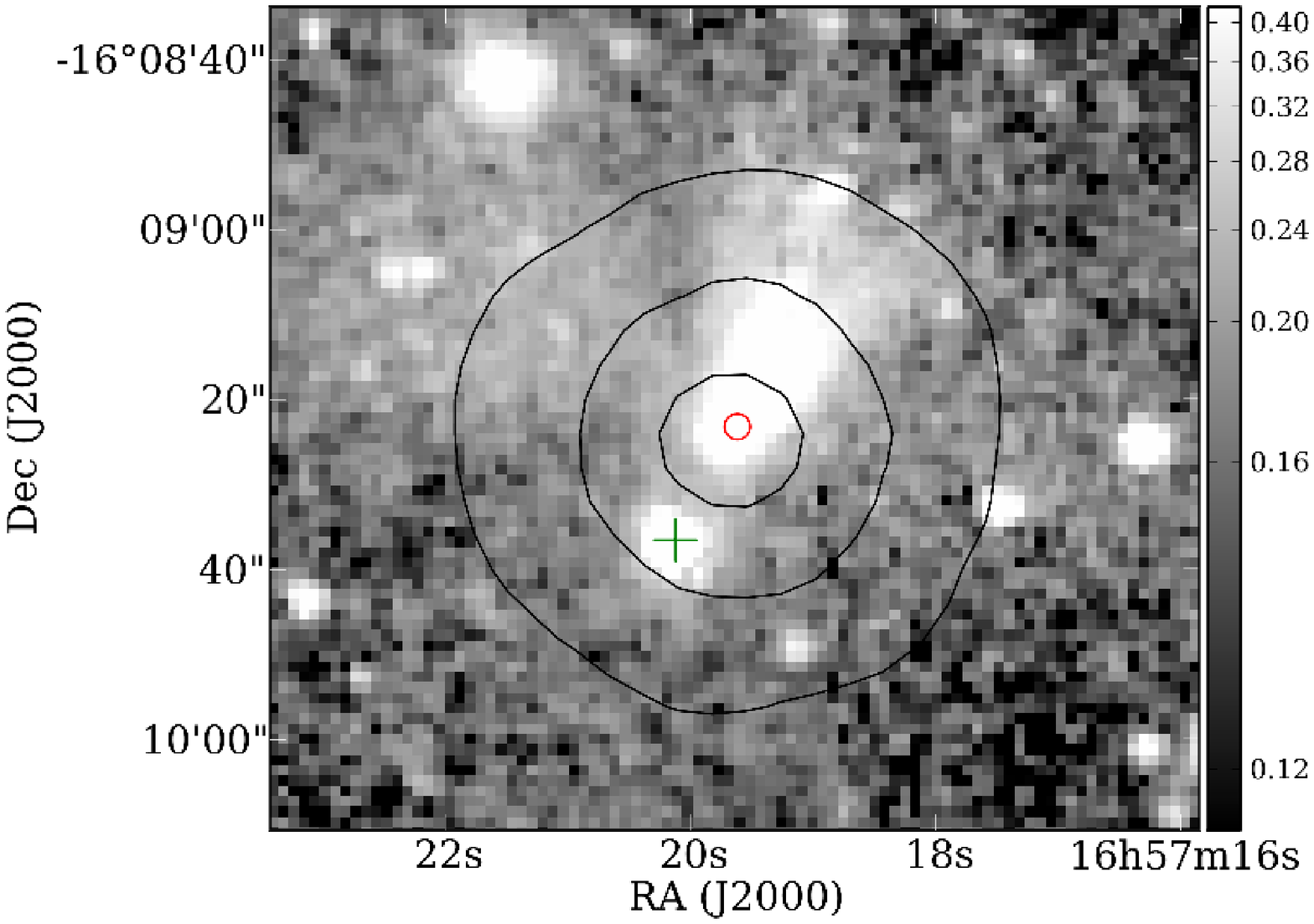}
\epsscale{0.65}
%\plottwo{PS/cb68_seds.eps}{PS/cb68_seds.eps}
%\plotone{PS/cb68_seds.eps}
\plotone{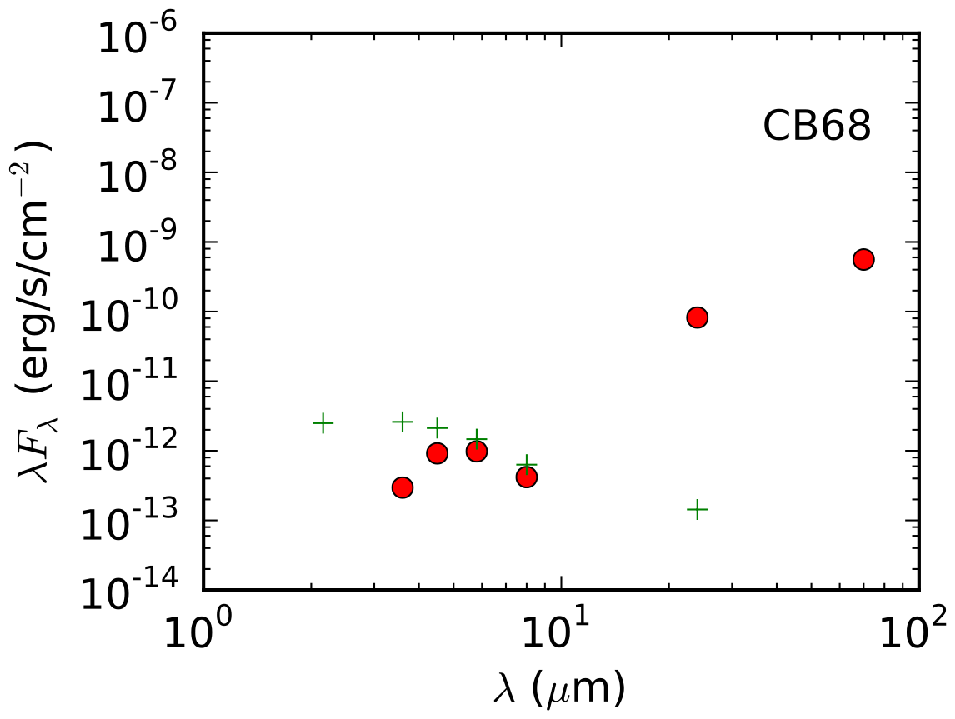}
\caption{Top: CB68 imaged by IRAC~2 (4.5\micron)  showing the reflection nebula extending towards the northwest in the direction of the outflow.  The black contours are MIPS~70\micron [50,100,500]~MJy/sterad.  Markers show the YSOc (circle, red in the online version) and southern star (cross, green in the online version) from the \Spitzer\ catalogue.  Bottom: CB68 SEDs for the YSO (circles) and the confusing star to the southeast (crosses).}
 \label{fig:cb68}
\end{figure}

CB68 has more than one associated source in the \Spitzer\ catalogue, as shown in Fig.~\ref{fig:cb68}.  The main source YSOc10 is strongly detected by IRAC and MIPS.  A very red SED, with no 2MASS detections and rising towards 100\micron\, confirms the Class~0 status (Fig.~\ref{fig:seds}).  To the southeast, $15''$ away, a second source is detected by IRAC.  This lies within the point spread of MIPS emission from the central source, and is bandfilled at 24\micron, and also shows some MUXBLEED artifacts to the north.  From its SED, this source is a star (Fig.~\ref{fig:cb68}) and though its MIPS flux is uncertain due to the confusion with IRAS~16544$-$1604 there is no evidence for a disk excess.  At $15''$ to the southeast, it has no associated 450\micron\ emission.  
%It is the reflection nebula and shocked H$_2$ emission associated with the southeast outflow lobe.  
All four IRAC bands show the outflow cavity extending to the northwest (Fig.~\ref{fig:cb68}).  There is also nebulosity to the west and east in the 3.6 and 4.5\micron\ bands.  

IRAS~16534$-$1557 lies fifteen arcminutes to the northwest of CB68.  This source is associated with \citet{lm99} core 224 or \citet{tachihara00a} core q2 (see Fig. 2 in \citealp{vallee07}) and has a rising IRAS spectrum characteristic of a YSO \citep{carballo92}, but lies beyond the range of our \Spitzer\ map.

% contours in cb68 with sources marked are 47, 100, 200
% plot SEDs for these two objects
% use readc2dcat to get this in Python

% Figure showing region observed
\clearpage

\clearpage

\clearpage

\clearpage

\clearpage

\clearpage

%\subsection{\Spitzer\ archival data}

\clearpage

\clearpage

\clearpage

\clearpage

\clearpage

\clearpage

\clearpage

\clearpage

\clearpage
\begin{deluxetable}{lccllll}
\tablecolumns{7}
\tabletypesize{\scriptsize}
\tablecaption{AORs in the \Spitzer\ Gould Belt Oph~N (Scorpius) Catalogue \label{tbl:aors}}
\tablewidth{0pt}
\tablehead{
\colhead{Instrument}&\colhead{RA(J2000)}
&\colhead{Dec(J2000)}&\colhead{PID}
&\colhead{Duration(mins)}&\colhead{Observation date/time}
&\colhead{AOR key}\\
}
\startdata
\multicolumn{7}{l}{\bf OphN~6}\\
IRAC    &16:21:24.19  &$-$19:56:01.80     &30574   &67.19   &2007-09-09 03:51:06.6     &19992320  \\ %IRAC_sco_1_1       sco_irac1            
IRAC    &16:21:24.48  &$-$19:56:05.80     &30574   &67.18   &2007-09-09 07:49:28.5     &19991808  \\ %IRAC_sco_1_2       sco_irac1            
MIPS    &16:21:29.60  &$-$19:57:45.00     &30574   &57.96   &2007-04-08 18:13:07.4     &20001280  \\ %MIPSC_sco_1_1      sco_mips1            
MIPS    &16:21:29.60  &$-$19:57:45.00     &30574   &57.96   &2007-04-09 00:39:24.0     &20000768  \\ %MIPSC_sco_1_2      sco_mips1            
      
\multicolumn{7}{l}{\bf OphN 5}\\
IRAC    &16:27:45.78  &$-$19:24:43.00     &30574   &25.38   &2006-09-21 13:19:07.2     &19960064  \\ %IRAC_sco_2_1        sco_irac2           
IRAC    &16:27:45.78  &$-$19:24:43.00     &30574   &25.38   &2006-09-21 17:17:44.2     &19959552  \\ %IRAC_sco_2_2        sco_irac2           
MIPS    &16:27:40.00  &$-$19:25:22.00     &30574   &30.83   &2007-04-08 10:55:48.0     &19968512  \\ %MIPSC_sco_2_1       sco_mips2           
MIPS    &16:27:40.00  &$-$19:25:22.00     &30574   &30.83   &2007-04-08 15:39:59.8     &19968000  \\ %MIPSC_sco_2_2       sco_mips2           
      
\multicolumn{7}{l}{\bf OphN 4}\\
IRAC    &16:31:53.68  &$-$19:35:03.00     &30574   &54.31   &2006-09-21 12:12:25.1     &19990784  \\ %IRAC_sco_3_1         sco_irac3          
IRAC    &16:31:53.97  &$-$19:35:07.00     &30574   &54.32   &2006-09-21 16:27:39.8     &19990272  \\ %IRAC_sco_3_2         sco_irac3          
MIPS    &16:32:21.50  &$-$19:43:36.10     &30574   &44.37   &2007-04-06 23:54:22.0     &19999744  \\ %MIPSC_sco_3_1        sco_mips3          
MIPS    &16:32:21.50  &$-$19:43:36.10     &30574   &44.37   &2007-04-07 05:44:06.0     &19998976  \\ %MIPSC_sco_3_2        sco_mips3          
      
\multicolumn{7}{l}{\bf OphN 3 (including c2d L158)}\\ % +L204?
IRAC    &16:50: 6.45  &$-$15:29:48.50     &30574   &68.54   &2006-09-26 15:36:32.5     &19959296  \\ %IRAC_sco_4_1               sco_irac4    
IRAC    &16:50: 6.83  &$-$15:29:47.80     &30574   &68.54   &2006-09-26 22:11:19.0     &19959040  \\ %IRAC_sco_4_2               sco_irac4    
MIPS    &16:49:60.00  &$-$15:30: 9.70     &30574   &78.19   &2007-04-07 20:15:50.3     &19966720  \\ %MIPSC_sco_4_1              sco_mips4    
MIPS    &16:49:60.00  &$-$15:30: 9.70     &30574   &78.19   &2007-04-08 04:12:25.9     &19966208  \\ %MIPSC_sco_4_2              sco_mips4    
IRAC    &16:49: 4.02  &$-$14:11:09.10     &30574   &48.44   &2006-09-26 20:52:06.4     &19990016  \\ %IRAC_sco_5_1               sco_irac5    
IRAC    &16:49: 4.40  &$-$14:11:08.40     &30574   &48.44   &2006-09-27 00:51:11.2     &19989504  \\ %IRAC_sco_5_2               sco_irac5    
MIPS    &16:48:53.60  &$-$14: 5:45.00     &30574   &78.19   &2007-04-08 19:09:17.5     &19996160  \\ %MIPSC_sco_5_1              sco_mips5    
MIPS    &16:48:53.60  &$-$14: 5:45.00     &30574   &78.19   &2007-04-09 01:35:34.1     &19995904  \\ %MIPSC_sco_5_2              sco_mips5    
IRAC    &16:47:08.00  &$-$13:58: 5.00     &139     &13.81   &2004-08-16 18:55:38.5     &5126400  \\ %IRAC - L158                L158         
IRAC    &16:47:08.00  &$-$13:57:55.00     &139     &13.81   &2004-08-17 07:35:56.5     &5126912  \\ %IRAC - L158 &$-$ 0001      L158            
MIPS    &16:47:15.60  &$-$13:59:28.80     &139     &20.46   &2005-03-10 02:36:55.6     &9415680  \\ %MIPS - L158                L158         
MIPS    &16:47:16.40  &$-$13:56:41.20     &139     &20.46   &2005-03-10 09:47:39.7     &9427456  \\ %MIPS - L158 - 0001         L158      
     
\multicolumn{7}{l}{\bf OphN 2 (including c2d L204C)}\\
IRAC    &16:47:32.48  &$-$12:06:21.40     &30574   &32.86   &2006-09-23 01:45:10.0     &19958784  \\ %IRAC_sco_6_1              sco_irac6     
IRAC    &16:47:32.85  &$-$12:06:20.60     &30574   &32.87   &2006-09-23 06:07:36.2     &19958528  \\ %IRAC_sco_6_2              sco_irac6     
MIPS    &16:47:33.80  &$-$12:09:06.40     &30574   &30.82   &2007-04-08 10:26:38.9     &19964160  \\ %MIPSC_sco_6_1             sco_mips6     
MIPS    &16:47:33.80  &$-$12:09:06.40     &30574   &30.82   &2007-04-08 15:10:50.8     &19963904  \\ %MIPSC_sco_6_2             sco_mips6     
IRAC    &16:47:39.00  &$-$12:21:09.00     &139     &8.49    &2004-08-17 07:47:01.5     &5127424 \\ %IRAC - L204C-2               L204C-2    
IRAC    &16:47:39.00  &$-$12:20:59.00     &139     &8.49    &2004-08-18 05:38:13.0     &5127936 \\ %IRAC - L204C-2 - 0001      L204C-2      
MIPS    &16:47:42.30  &$-$12:21:11.70     &139     &20.39   &2005-03-10 02:18:26.2     &9408512 \\ %MIPS - L204C-2               L204C-2    
MIPS    &16:47:35.70  &$-$12:21:06.30     &139     &20.39   &2005-03-10 09:29:20.8     &9422080 \\ %MIPS - L204C-2 - 0001        L204C-2    
      
\multicolumn{7}{l}{\bf OphN 1}\\
IRAC    &16:47:24.26   &$-$9:46:39.40     &30574   &43.07   &2006-09-23 18:42:25.8     &19988224  \\ %IRAC_sco_7_1             sco_irac7      
IRAC    &16:47:24.63   &$-$9:46:38.60     &30574   &43.07   &2006-09-24 01:31:23.5     &19987712  \\ %IRAC_sco_7_2             sco_irac7      
MIPS    &16:47:24.01   &$-$9:52:44.60     &30574   &30.82   &2007-04-08 17:43:41.6     &19994368  \\ %MIPSC_sco_7_1            sco_mips7      
MIPS    &16:47:24.01   &$-$9:52:44.60     &30574   &30.82   &2007-04-08 23:11:15.0     &19994112  \\ %MIPSC_sco_7_2            sco_mips7      

\multicolumn{7}{l}{\bf L234E}\\
IRAC    &16:48:09.00  &$-$10:55:56.00     &139     &12.05   &2004-08-17 07:53:06.1     &5128448  \\%IRAC - L234E          L234E                
IRAC    &16:48:09.00  &$-$10:55:46.00     &139     &12.04   &2004-08-18 05:44:20.5     &5128960  \\%IRAC - L234E - 0001   L234E                
IRAC    &16:48:12.20  &$-$10:55:59.60     &139     &20.39   &2005-03-10 02:00:09.4     &9411072  \\%MIPS - L234E          L234E                
IRAC    &16:48:05.80  &$-$10:55:52.40     &139     &20.39   &2005-03-10 09:10:57.7     &9431808  \\%MIPS - L234E - 0001   L234E          

%\multicolumn{7}{l}{\bf L260}\\
% IRAC    &16:47:07.70   &$-$9:35:56.00     &139     &8.81    &2004-08-17 08:02:26.7     &5125632  \\ %IRAC - L260          L260          
% IRAC    &16:47:07.70   &$-$9:35:46.00     &139     &8.81    &2004-08-18 05:53:39.0     &5126144  \\ %IRAC - L260 - 0001   L260          
% MIPS    &16:47:10.90   &$-$9:35:59.40     &139     &18.79   &2005-03-10 01:43:24.9     &9430016  \\ %MIPS - L260         L260           
% MIPS    &16:47:04.50   &$-$9:35:52.60     &139     &18.79   &2005-03-10 08:54:06.9     &9429760  \\ %MIPS - L260 - 0001  L260           
      
\multicolumn{7}{l}{\bf L146 / CB68}\\
IRAC    &16:57:20.50  &$-$16:09:02.00     &139     &8.83    &2004-08-17 08:23:08.3     &5130752  \\ %IRAC - CB68           CB68         
IRAC    &16:57:20.50  &$-$16:08:52.00     &139     &8.83    &2004-08-18 07:55:40.2     &5131264  \\ %IRAC - CB68 - 0001    CB68         
MIPS    &16:57:20.10  &$-$16:10:25.80     &139     &18.87   &2005-04-04 05:55:18.0     &9425408  \\ %MIPS - CB68          CB68          
MIPS    &16:57:20.90  &$-$16:07:38.20     &139     &18.88   &2005-04-05 22:55:58.5     &9439232  \\ %MIPS - CB68 - 0001   CB68          

%IRAC         &16:34:27.00  &$-$15:46:48.00         &139    11.68   2004-08-16 18:45:55.5      5121280  \\ %IRAC - L43           L43           
%IRAC         &16:34:27.00  &$-$15:46:38.00         &139    11.67   2004-08-17 07:26:21.7      5121792  \\ %IRAC - L43 - 0001    L43           
%MIPS         &16:34:26.40  &$-$15:48:11.50         &139   18.86   2005-03-10 01:05:35.8      9409792  \\ %MIPS - L43            L43          
%MIPS         &16:34:27.60  &$-$15:45:24.50         &139   18.86   2005-03-10 10:06:38.5      9433344  \\ %MIPS - L43 - 0001     L43          
\enddata
\tablecomments{The original AOR names of regions OphN~1--6 were Sco~1--6.}
\end{deluxetable}

\clearpage
\begin{deluxetable}{llp{4cm}}
\tablewidth{0pc}
\tablecaption{Associations with Lynds Dark Nebulae and molecular cores. \label{tbl:assc}}
\tablehead {
\colhead{Region}     &\colhead{CO\tablenotemark{a}}  &\colhead{LDN  (opacity) \tablenotemark{b} }\\
}
{}
\startdata
OphN~1   &U, u1,u2     &L260 (6), L255 (6)\\
L234E   &T, t             &L234 (3)\\
OphN~2   &S, s1-5       &L204 (6) \\   
OphN~3a     &R, r1,r2        &L152 (3), L158 (6), L162(6), L163(2), L141 (5),L137 (4)\\
OphN~3b     &P, p1,p2       &--\\
CB68    &Q, q1           &L146 (5)\\
OphN~4          &C,--                 &L1757 (6)\\
OphN~5   &C,--                 &L1752 (4)\tablenotemark{c}\\
OphN~6   &A, B, a, b      &L1719 (5)\tablenotemark{c}\\
\enddata
\tablenotetext{a}{Cores identified from $^{13}$CO/C$^{18}$O \protect\citep{tachihara00a}}
\tablenotetext{b}{{Lynds Dark Nebula} \protect\citep{lynds62}}
\tablenotetext{c}{Dark Nebulae L1719 and L1752 lie in the constellation of Scorpius, whereas all the others listed here are within Ophiuchus.}
%C$^{18}$O cloud and core designations come from \protect\citet{tachihara00a} and Lynds dark clouds (LDC) from %\protect\citet{lynds62}. 
\end{deluxetable}

\clearpage
\begin{deluxetable}{@{}ll@{}}
\tabletypesize{\scriptsize}
\tablewidth{0pc}
\tablecaption{Detection statistics with IRAC and 2MASS.  \label{tbl:detections}}
% see 20090407.html
\tablehead{
% From 20090429 data inc. CB68 and L234E
\colhead{Detection type}         &\colhead{Detections}
}
{}
\startdata
% sn = 3
%IDL> .r c2d_detect.pro
%% Compiled module: $MAIN$.
%Detections in...
At least one IRAC band: &106200\\
All 4 bands : &5378\\
Only 3 bands : &7902\\
Only 2 bands : &50632\\
Only 1 bands : &42288\\
&\\
  MIPS 1               &845\\
  MIPS 1 and IRAC      &762\\
  MIPS and all 4 IRAC bands &449\\
&\\
  2MASS k               &10919\\
  2MASS k and h         &10822\\
  2MASS k and at least one IRAC band &10917\\
  2MASS k and all 4 IRAC bands &4723\\
  2MASS k but not IRAC &2\\
  IRAC band 1 but not 2MASS k    &89038\\
&\\
  MIPS, all 4 IRAC bands, and 2MASS k &196\\
\enddata
% Per IRAC band:
% S/N of 3                136853       94166       18320        8821
% S/N of 5               101913       64178        9618        5465
% S/N of 10                50565       28830        4742        2696
% S/N of 15                23182       13459        2446        1460
% Excluding extended       70734       45884        6930        3859
% 2MASS ass.                5208        5238        5039        3094
% A or B in all and not extended        3291
% A or B in all                         3339
% 2MASS and A or B in four              3039
% 2MASS and A or B in IRAC1+2           5189
\tablecomments{Detections are counted above a signal-to-noise ratio of 3 and in the overlap area of all 4 IRAC bands only.}
\end{deluxetable}

\clearpage
\begin{deluxetable}{lrrrr}
%\tabletypesize{\scriptsize}
\tablecolumns{5}
\tablewidth{0pc}
\tablecaption{Detection of sources above various S/N thresholds in individual IRAC bands. \label{tbl:sn}}
%\begin{tabular}{@{}lrrrr@{}}
% also from 20090429 inc. CB68 and L234E
\tablehead{
\colhead{S/N}&\colhead{3.6\micron }&\colhead{4.5\micron }&\colhead{5.8\micron }&\colhead{8.0\micron} \\
}
{}
\startdata
S/N of 3        &136853     &94166       &18320      &821\\
S/N of 5        &101913     &64178       &9618       &5465 \\
S/N of 10      &50565       &28830       &4742       &2696\\
S/N of 15\quad      &23182       &13459       &2446       &1460\\
%Excluding extended    &   67450   &    43695    &     6610    &    3663\\
%2MASS ass.            &    4954   &     4982    &     4793    &    2931\\
\tablecomments{Sources are counted here even if they fall outside the 4-band overlap area, so counts can be larger than in Table~\protect\ref{tbl:detections}.}
\enddata
\end{deluxetable}

\clearpage
\begin{deluxetable}{r l l l l l l l l}
\tabletypesize{\scriptsize}
\rotate
\tablewidth{0pt}
\tablecaption{YSO candidates with IRAC detections \label{tbl:ysos}}
\tablehead{
%\begin{tabular}{r l l l l l l l}
% RA/Dec from sco_all_YSOs_new_txtcoords.txt.  Regions from Sco n_YSOs.txt.  IDs from SIMBAD with file sco_all_YSOs_new_radec.txt
\colhead{YSOc}&\colhead{SSTgbs  }&\colhead{Field }&\colhead{Class }&\colhead{$\alpha$\protect\tablenotemark{a}}&
\colhead{ID (offset) }&\colhead{Submm core }&\colhead{References}\\}
{}
\startdata
1 &J164658.3$-$093519   &OphN1 &I &$0.66$ &IRAS~16442-0930 ($0.9'$) &L260 SMM1 &1,2,3 \\%CWW92,BATC96,VRC01 \\ % De Grijp et al. 87 AGN candidate
2 &J164817.6$-$141109 &OphN3 &II &$-0.70$ &V2507 Oph($9.0''$), Oph4 ($9.0''$) \tablenotemark{b} & &4,5\\%D05,AM94\\
3 &J164821.9$-$141043  &OphN3 &II &$-1.03$ &IRAS~16455-1405 (0.27''), PDS~91  & &1,6\\%CWW92,RZ93\\
% & 	RC02,BATC96 (Class I),CWW92\\ %L158, see Visser 2002
4 &J164845.6$-$141636  &OphN3 &II &$-0.97$ &IRAS~16459-1411, V2508 Oph, Oph6 (0.16'') &L162 SMM1&5,7,8\\%AM94,A09,V02\\
% & 	lso V* V2508 Oph (0.16''), WaOph6
5 &J164900.8$-$141711  &OphN3 &II &$-1.28$ &Oph~5 (0.12''), HBC~654& &5\\%AM94\\
6\tablenotemark{c} &J164905.6$-$153713  &OphN3 &III &$-1.87$ &IRAS~16462-1532 (7.45'') &\\
7\tablenotemark{c} &J162137.7$-$200037  &OphN6 &III &$-2.09$ &--&\\
8 &J162204.3$-$194327 &OphN6 &F &$0.02$ &IRAS~16191-1936 (1.04'')&L1719B &1,2,9\\%CWW92,BATC96,K05\\
9 &J162209.6$-$195301  &OphN6 &III &$-2.11$ &--&\\
10&J165719.6$-$160923  &CB68 &I &$2.40$ &IRAS~16544-1604 (12.05'')&CB68,CB68SMM1 &10,1,11\\%CB88, CWW92,HSW99 \\
\enddata
\tablenotetext{a}{\jh{Infrared spectral index $\alpha$ from a fit to fluxes from $K_S$ band to MIPS~24\micron\ \protect\citep{c2ddel}.}}
\tablenotetext{b}{The Spitzer detection and V2507~Oph are separate stars, probably in a binary.  See Appendix.~\ref{sect:sco3}.}
\tablenotetext{c}{Probably an AGB star.  See Sect.~\ref{sect:agb}.}
\tablecomments{References: (1) \citet{carballo92}; (2) \citet{batc96}; (3) \citet{visser01}; (4) \citet{ducourant05}; (5) \citet{am94}; (6) \citet{reipurth93}; (7) \citet{andrews09}; (8) \citet{visser02}; (9) \citet{kirk05}; (10) \citet{cb88}; (11)\citet{huard99} . }
\end{deluxetable}

\clearpage
\begin{deluxetable}{l c c c c c c c c c c}
\tabletypesize{\scriptsize}
\rotate
\tablecolumns{11}
\tablewidth{0pc}
\tablecaption{Fluxes for YSO candidates with IRAC detections. \label{tbl:fluxes}}
%\begin{tabular}{l l l c c c c c c c c c c}

\tablehead{
\colhead{}&\multicolumn{3}{c}{\hrulefill \quad2MASS\quad \hrulefill}&\multicolumn{4}{c}{ \hrulefill\quad \Spitzer\ IRAC \quad \hrulefill }&
\multicolumn{3}{c}{\hrulefill\quad \Spitzer\ MIPS\quad \hrulefill}\\

\colhead{YSOc }&\colhead{J }&\colhead{H }&\colhead{ K$_\mathrm{s}$ }&\colhead{F3.6 }&\colhead{F4.5 }&\colhead{F5.8 }&\colhead{F8.0 }&\colhead{F24 }&\colhead{F70 }&\colhead{F160}  \\
\colhead{}&\colhead{mJy }&\colhead{ mJy }&\colhead{ mJy }&\colhead{ mJy }&\colhead{mJy }&\colhead{mJy }&\colhead{mJy }&\colhead{mJy }&\colhead{mJy }&\colhead{mJy} \\
}
{}
\startdata
1&$1.8\pm 0.1$ & $8.3\pm 0.3$ &$26.8\pm 0.8$&81.8$\pm$ 4.3 &150.0$\pm$ 7.8 & 231$\pm$  11 & 189$\pm$  12 &2430$\pm$ 248 &   --\tablenotemark{a}  &3700$\pm$   730   \\%        16465826-0  N0 N0 off MIPS area             YSOc\_red 
 2&$ 50.9\pm 1.7$ &$90.4\pm 4.4 $&$105.0\pm 2.6$         &71.4$\pm$ 3.5 &64.6$\pm$ 3.1 &  68$\pm$   3 &  90$\pm$   4 & 168$\pm$  16 &  87$\pm$  12 &   $<2200$           \\%    16481763-1 A7 N0 YSOc\_star+dust
3&$123.0\pm 3.2$ &$298.0\pm 7.7$ &$477.0\pm 7.9$    &641.0$\pm$31.9 &559.0$\pm$30.5 & 539$\pm$  26 & 471$\pm$  23 & 639$\pm$  59 & 476$\pm$  49 &  $<2200$   \\% 16482187-1 A1 N0 YSOc\_star+dust
4&$519.0\pm11.5$ &$957.0\pm33.5$ &$1200.0\pm29.9$     &1010.0$\pm$51.5 &889.0$\pm$46.0 & 816$\pm$  40 & 925$\pm$  44 &1310$\pm$ 124 & 873$\pm$  92 &  $<2200$  \\%   16484562-1 A1 N0  YSOc\_star+dust
5&$371.0\pm 8.2$ &$571.0\pm21.0$ &$575.0\pm11.1$      &390.0$\pm$19.9 &315.0$\pm$19.1 & 274$\pm$  13 & 270$\pm$  13 & 303$\pm$  28 &  92$\pm$  14 &$<2200$          \\%&   0$\pm$   0      16490082-1 B7 N0 YSOc\_star+dust
6&$140.0\pm 3.2$ &663.0$\pm$25.6 &1120.0$\pm$21.8       &1000.0$\pm$51.6 &778.0$\pm$43.4 & 742$\pm$  36 & 517$\pm$  25 & 146$\pm$  14 &   $<11$ &$<2200$\\%&  YSOc\_star+dust 0$\pm$   0     16490559-1 U0 N0 (70 micron ND.  70 micron never bandfilled.)
7&437.0$\pm$ 8.4 &1150.0$\pm$38.2 &1510.0$\pm$25.0          &1100.0$\pm$57.9 &809.0$\pm$40.4 & 694$\pm$  34 & 472$\pm$  22 & 114$\pm$  11 &   $<11$ &$<2200$\\%&   0$\pm$   0     16213768-2 U0 N0 (ND) YSOc\_star+dust
8&27.8$\pm$ 0.7 &136.0$\pm$ 2.9 &347.0$\pm$ 8.3                &759.0$\pm$38.3 &1090.0$\pm$56.3 &1440$\pm$  69 &2010$\pm$ 110 &3730$\pm$ 390 &5320$\pm$ 573 &   8700$\pm$   730        \\%  16220435-1 A1 N0  good 70 YSOc           
9&144.0$\pm$ 3.8 &208.0$\pm$ 4.8 &184.0$\pm$ 4.1          &109.0$\pm$ 6.6 &71.5$\pm$ 4.1 &  50$\pm$   3 &  33$\pm$   2 &  16$\pm$   2 &   $<11$ &$<2200$\\%&   0$\pm$   0       16220961-1 U0 N0 (ND) YSOc\_star+dust
10 & 0 & 0 & 0 & 0.4$\pm$ 0.0 & 1.4$\pm$ 0.1 &1.9$\pm$0.1 &1.1$\pm$0.1 &659$\pm$61 &13100$\pm$1210 &-- \\% CB68 YSOc\_red
%Galaxy&          YSOc\_PAH-e & 2.0$\pm$ 0.2 & 1.4$\pm$ 0.2 &   5$\pm$   1 &  42$\pm$   2 &  23$\pm$   2 & 455$\pm$  50 &   0$\pm$   0\\%            16482880-1 
\enddata
\tablenotetext{a}{Source lies outside the 70\micron\ map.}

%  The letters in brackets indicate whether the non-detection is because the sources is outside the MIPS region (N) or bandfilled (U).  Imtype =0 means too close to the edge of the frame to have a reliable fit. 

\end{deluxetable}

\clearpage
\begin{deluxetable}{r c c c c l}
\tabletypesize{\scriptsize}
\rotate
\tablecolumns{6}
\tablewidth{0pc}
\tablecaption{YSO candidates identified from MIPS 24 micron  and 2MASS in regions with no IRAC coverage.\label{tbl:mipsysocs}}
\tablehead{
\colhead{YSOc }&\colhead{SSTgbs }&\colhead{Field }&\colhead{Class }&\colhead{ID(offset) }&\colhead{References}\\
}
{}
\startdata
%#kmips YSOs from newkmipsysos.cat converted to text format
%M1& 16:47:41.4& -14:10:07.9& Sco3 &F  &13.8978 &6.49038 &1.92941 & \\
%M2&16:48:42.5& -15:29:31.4& Sco3 & I & 13.9647 &7.49555 &2.45022 &\\ 
%M3&16:50:28.0& -15:05:33.1& Sco3 &F  & 13.9277 &5.70315 &1.60328 &\\ 
%M4&16:50:51.2& -14:44:07.7& Sco3 &F  & 13.9773 &6.60804 &--    & \\ 
M1\tablenotemark{a}& J165121.4$-$160259& OphN3 &III&IRAS~16484$-$1557 ($7''$), ASAS 165122-1602.9 ($5.9''$) &1 \\%PPS05 \\          %&6.45223 &4.03226 &--     \\ 
%M6& 16:31:39.8& -20:20:44.4& OphN4 &F  &13.0822 &5.0363 &0.609608 &6dFGS gJ163139.8$-$202044 ($0.45''$)\\ 
M2\tablenotemark{a}& J163142.5$-$203814& OphN4 &II &\\  %&8.63216 &4.93786 &3.26878 &--\\ 
%M8& 16:31:58.4& -20:06:24.1& OphN4 &II &13.4774 &8.21745 &--    &-- \\ 
%M9& 16:28:28.2& -19:00:37.2& OphN5 &II &13.6461 &8.96715 &--    &--\\ 
M3& J162214.8$-$204540& OphN6 &II &\\  %&11.2237 &7.61936 &--    &-- \\ 
M4& J162229.8$-$200248& OphN6 &II  &2RXP~J162230.1-200256 (9.1'') &\\  % &8.59581 &4.53526 &2.07326&-- \\  
M5\tablenotemark{b}&J164915.3$-$142209 & OphN3 &II &V1121 Oph($0.08''$) &2,3\\%AM94,CWW92\\

% RA/Dec format
% M1& 16:51:21.4& -16:02:58.8& OphN3 &III&6.45223 &4.03226 &--     &IRAS~16484$-$1557 ($7''$)\\ 
% %M6& 16:31:39.8& -20:20:44.4& OphN4 &F  &13.0822 &5.0363 &0.609608 &6dFGS gJ163139.8$-$202044 ($0.45''$)\\ 
% M2& 16:31:42.5& -20:38:14.3& OphN4 &II &8.63216 &4.93786 &3.26878 &--\\ 
% %M8& 16:31:58.4& -20:06:24.1& OphN4 &II &13.4774 &8.21745 &--    &-- \\ 
% %M9& 16:28:28.2& -19:00:37.2& OphN5 &II &13.6461 &8.96715 &--    &--\\ 
% M3& 16:22:14.8& -20:45:39.9& OphN6 &II &11.2237 &7.61936 &--    &-- \\ 
% M4& 16:22:29.8& -20:02:47.7& OphN6 &II &8.59581 &4.53526 &2.07326&-- \\  
\enddata
\tablenotetext{a}{Probably an AGB star.  See Sect.~\ref{sect:agb}.}
\tablenotetext{b}{V1121 Oph lies outside the IRAC mapping region at 4.5 and 8.0\micron\ so is not classified as a YSOc by the c2d pipeline.}
\tablecomments{References:  (1) \citet{pojmanski05}; (2) \citet{am94}; (3)\citet{carballo92}.} 
\end{deluxetable}

\clearpage
\begin{deluxetable}{l c c c c c c c c c c}
\tabletypesize{\scriptsize}
\rotate
\tablecolumns{11}
\tablewidth{0pc}
\tablecaption{Fluxes for YSO candidates identified from MIPS 24 micron and 2MASS. \label{tbl:mipsfluxes}}
\tablehead{
\colhead{}&\multicolumn{3}{c}{\hrulefill \quad2MASS\quad \hrulefill}&\multicolumn{4}{c}{ \hrulefill\quad WISE\quad \hrulefill }&
\multicolumn{3}{c}{\hrulefill\quad \Spitzer\ MIPS\quad \hrulefill}\\
\colhead{YSOc }&\colhead{J }&\colhead{ H }&\colhead{ K$_S$ }&\colhead{F3.4 }&
\colhead{F4.6 }&\colhead{F12}&\colhead{F22 }&\colhead{F24 }&\colhead{F70 }&
\colhead{ F160}\\ 
\colhead{}&\colhead{mJy }&\colhead{ mJy }&\colhead{ mJy }&\colhead{mJy }&
\colhead{mJy }&\colhead{mJy }&\colhead{ mJy} &\colhead{mJy }&\colhead{mJy }&
\colhead{mJy}\\  
}
{}
\startdata
M1 &$969.0\pm18.7$ &$1550.0\pm28.5$ &$1750.0\pm33.9$ &$1029.7\pm38.9$ &$1086.9\pm24.0$ &$406.91\pm7.12$ &$185.34\pm4.27$&$178\pm17$ &$<11$ &$<2200$\\  %&6.45223 &4.03226 &--     \\ 
M2 &$93.1\pm 1.8$ &$203.0\pm 3.9$ &$235.0\pm 4.6$ &$202.6\pm 2.4$ &$154.7\pm 2.4$ &$158.31\pm2.62$ &$93.06\pm2.40$ &$77\pm7$ &$40\pm7$ &$<2200$\\  %&8.63216 &4.93786 &3.26878 &--\\ 
M3 &$21.8\pm 0.5$ &$24.0\pm 0.6$ &$21.6\pm 0.4$ &$11.8\pm 0.3$ &$ 8.5\pm 0.2$ &$4.79\pm0.21$ &$6.92\pm1.21$ &$7\pm1 $&$<11$ &$<2200$\\  %&11.2237 &7.61936 &--    &-- \\ 
M4 &$115.0\pm 2.3 $&$215.0\pm 4.2$ &$243.0\pm 4.3$ &$222.2\pm 4.9$ &$194.2\pm 3.6$ &$115.75\pm1.92$ &$128.10\pm2.95$&$112\pm10$&$120\pm16$ &$<2200$\\  % &8.59581 &4.53526 &2.07326&-- \\  
M5&$762.0\pm27.4$ &$1070.0\pm23.6$ &$1100.0\pm26.2$ &$746.6\pm22.7$ &$744.4\pm15.1$ &$3170.32\pm73.00$ &$5915.10\pm49.03$ & $3850\pm221$\tablenotemark{a}&$2400\pm 79$\tablenotemark{a} &$2050\pm730$\\
\enddata
\tablenotetext{a}{24 and 70\micron\ fluxes from $100''$ radius aperture photometry.  The listed c2d pipeline point-source 24\micron\ flux was $410\pm200$ mJy. }
%\end{landscape}
\end{deluxetable}

\clearpage
\begin{deluxetable}{l c c c}
\tablewidth{0pc}
\tablecaption{Core masses above extinction thresholds. \label{tbl:extinctionmasses}}
\tablehead{
\colhead{Region} &\colhead{$M(A_V > 6)$\tablenotemark{a}}&\colhead{$M(A_V > 10)$} &\colhead{$M(A_V > 20)$} \\ 
\colhead{} &\Msun&\Msun&\Msun \\ 
}
{}
\startdata
% on 150" maps / 120" maps(CB68/L234E)
OphN~1 &30  &10 &1\\
OphN~2 &51  &17 &2\\
OphN~3 &149 &52 &7\\
OphN~4 &9 &0 &0\\
OphN~5 &6  &0 &0\\    
OphN~6 &40  &5 &0\\
CB68 &1 &0 &0\\
L234E &8 &1 &0 \\
   &{\bf 294} &{\bf 85} &{\bf 10}\\
\enddata
\tablenotetext{a}{The $A_V=6$ masses are lower limits as these regions extend beyond the limits of the map.  See Fig.~\protect\ref{fig:extinction}.} 
\end{deluxetable}

\clearpage
\begin{deluxetable}{l c c}
\tablewidth{0pc}
\tablecaption{100/160 micron spectral slope and average dust temperatures \label{tbl:tdust}}
\tablehead{
\colhead{Region}   &\colhead{slope\tablenotemark{a}} &\colhead{$T_\mathrm{d}$/K\tablenotemark{b}$/\beta = 2.0$ }
}
{}
\startdata 
OphN~1    &0.35    & 15.6         \\
OphN~2    &0.39     &16.1         \\
OphN~3S  &0.37     &15.9         \\
OphN~3N  &0.40     &16.3        \\
OphN~4    &0.39     &16.1        \\
OphN~5    &0.45     &17.0        \\
OphN~6    &0.44     &16.8        \\
\enddata
\tablenotetext{a}{Uncertainty in slope is $\pm0.002$. Estimated systematic uncertainty of 10 percent leads to $\pm 0.4$~K absolute temperature uncertainty.}
\tablenotetext{b}{Color-corrected dust temperature for $\beta = 2.0$. Changing from $\beta = 2.0$ to $\beta = 1.7$ (appropriate for \citealp{oh94} model OH5 dust opacity) leads to dust temperatures that are $0.8$~K systematically higher.}
\end{deluxetable}

\clearpage
\begin{deluxetable}{l l l l l}
\tablewidth{0pc}
\tablecaption{Parallaxes and proper motions for early-type stars in OphN~6. \label{tbl:bstars}}
\tablehead{
\colhead{  HIP }&\colhead{Spectral type }&\colhead{Parallax }&\colhead{$mu_{\alpha}$ }&\colhead{ $\mu_{\delta}$}\\
\colhead{        }&\colhead{             }&\colhead{mas   }&\colhead{arcsec/yr  }&\colhead{arcsec/yr} \\
}
{}
\startdata
80019 & B9\tablenotemark{a}  &$5.17\pm1.10$&$-12.01\pm1.38$ &$-25.54\pm1.01$\\
80024 & B8\tablenotemark{b} &$6.12\pm0.78 $&$-11.89\pm1.01$ &$-24.25\pm0.71$\\
80062 & B9.5 Va\tablenotemark{c}       &$9.82\pm2.44$ &$ -17.31\pm2.88$ &$-19.40\pm2.08$\\
80063 &  A0\tablenotemark{d}        &$7.45\pm3.42$ &$-17.55\pm 4.17$ &$-13.24\pm2.86$\\
\enddata
\tablecomments{Parallaxes and proper motions are taken from \citet{vanleeuwen07}. }
\tablenotetext{a}{\citet{hernandez05}}
\tablenotetext{b}{Strictly, kB8HeA0hB8IImA2IbSi(Cr:) \citep{garrison94}.} 
% \citet{hernandez05} give the type as A8, but this is almost certainly a typographical error.}
\tablenotetext{c}{HIP 80062/63 form a visual double \citep{gray06}}
\tablenotetext{d}{kA0hA0VmA2 \citep{abt79}}
\end{deluxetable}

\label{lastpage}
\end{document}